\documentclass[prx, aps, twocolumn, amsmath, amssymb, superscriptaddress]{revtex4-2}
\usepackage[english]{babel}
\usepackage{graphicx,verbatim}
\usepackage{braket}
\usepackage{color}

\usepackage{soul}
\usepackage[x11names]{xcolor}

\usepackage[
    colorlinks=true,
    linkcolor=RoyalBlue4,
    citecolor=RoyalBlue4,
    urlcolor=RoyalBlue4,
    breaklinks=true
]{hyperref}

\begin{document}

\title{Compact representation and long-time extrapolation of real-time data for quantum systems using the ESPRIT algorithm}

\author{Andr\'e Erpenbeck}
 \affiliation{Department of Physics, University of Michigan, Ann Arbor, Michigan 48109, USA}
 \affiliation{Department of Physics and Astronomy, University of Georgia, Athens, GA 30602, USA}
 \affiliation{ Center for Simulational Physics, University of Georgia, Athens, GA 30602, USA}
\author{Yuanran Zhu}
 \affiliation{Applied Mathematics and Computational Research Division, Lawerence Berkeley National Laboratory, Berkeley, CA 94720, USA}
\author{Yang Yu}
 \affiliation{Department of Physics, University of Michigan, Ann Arbor, Michigan 48109, USA}
\author{Lei Zhang}
 \affiliation{Department of Physics, University of Michigan, Ann Arbor, Michigan 48109, USA}
 \author{Richard Gerum}
 \affiliation{Department of Physics and Astronomy, York University, Toronto, Ontario, Canada}
\author{Olga Goulko}
 \affiliation{Department of Physics, University of Massachusetts Boston, Boston, Massachusetts 02125, USA}
\author{Chao Yang}
 \affiliation{Applied Mathematics and Computational Research Division, Lawerence Berkeley National Laboratory, Berkeley, CA 94720, USA}
\author{Guy Cohen}
 \affiliation{The Raymond and Beverley Sackler Center for Computational Molecular and Materials Science, Tel Aviv University, Tel Aviv 6997801, Israel}
 \affiliation{School of Chemistry, Tel Aviv University, Tel Aviv 6997801, Israel}
\author{Emanuel Gull}
 \affiliation{Department of Physics, University of Michigan, Ann Arbor, Michigan 48109, USA}
 \affiliation{Institute of Theoretical Physics, Faculty of Physics, University of Warsaw, Warsaw, Poland}

\date{\today}

\begin{abstract}

    Representing real-time data as a sum of complex exponentials provides a compact form that enables both denoising and extrapolation. 
    As a fully data-driven method, the Estimation of Signal Parameters via Rotational Invariance Techniques (ESPRIT) algorithm is agnostic to the underlying physical equations, making it broadly applicable to various observables and experimental or numerical setups. 
    In this work, we consider applications of the ESPRIT algorithm primarily to extend real-time dynamical data from simulations of quantum systems.
    We evaluate ESPRIT's performance in the presence of noise and compare it to other extrapolation methods. 
    We demonstrate its ability to extract information from short-time dynamics to reliably predict long-time behavior and determine the minimum time interval required for accurate results. 
    We discuss how this insight can be leveraged in numerical methods that propagate quantum systems in time, and show how ESPRIT can predict infinite-time values of dynamical observables, offering a purely data-driven approach to characterizing quantum phases.
    
\end{abstract}

\maketitle

\section{Introduction}

    Real-time dynamics plays a central role across physics, including condensed matter, optics, nuclear physics, high-energy, and quantum computing. 
    Compact representations of real-time quantum data help assess the information content of the dynamics, support denoising, and enable reliable extrapolation to longer times. 
    This is crucial in numerical simulations, where the computational costs grow rapidly with time and precision, and in experiments, where data is often noisy and limited in time. 
    Structured representations also aid post-processing tasks such as Fourier transforms or analytic continuation, help distinguish physical features from noise or artifacts, and improve efficiency through data compression.
    
    Compact representations have found broad use in many areas of physics. 
    In field theories, where Green's functions are central to characterizing system properties, a variety of strategies have been developed.
    For imaginary-time or Matsubara Green's functions, representations have been evaluated from uniform power mesh grids \cite{Ku_Band_2002}, to spline interpolation \cite{kananenka_efficient_2016_2}, and expansions in Legendre \cite{Boehnke_Orthogonal_2011, kananenka_efficient_2016} or Chebyshev polynomials \cite{Gull_Chebyshev_2018}. 
    More recent advances include the intermediate representation (IR) \cite{Shinaoka_Compressing_2017, Shinaoka_Overcomplete_2018, Chikano_Performance_2018}, the discrete Lehmann representation (DLR) \cite{Kaye_Discrete_2022}, discrete pole representations for Feynman diagrams \cite{Gazizova_Feynman_2024}, and representations based on sums over complex poles \cite{Ying_pole_2022_a,Ying_pole_2022_b,zhang_minimal_matsubara_2024, zhang_minimal_matrix_2024}, which are particularly well suited for analytical continuation.
    Early work on calculating complex pole representations traces back to Prony's method \cite{Prony_1795, Beylkin_approximation_2005, Beylkin_approximation_2010}, and the field has since expanded with a range of modern algorithms developed within the applied mathematics community. 
    Recent efforts aim to identify minimal sets of poles for accurate and efficient representations \cite{zhang_minimal_matsubara_2024, zhang_minimal_matrix_2024}.
    
    These ideas naturally extend to real-time Green's functions, where poles correspond to decaying complex exponentials -- a structure that aligns well with quantum dynamics.
    While the estimation of complex exponentials from time-series data is well studied in signal processing \cite{sarkar_using_1995, hua_matrix_1990, Beylkin_approximation_2010, roy_esprit_1989}, it is equally relevant to quantum systems. 
    Identifying minimal exponential representations remains an active area of research \cite{zhang_minimal_2025}, as it enables both denoising through physical constraints \cite{Kemper_Denoising_2024,Yu_Denoising_2024} and data compression \cite{Dong_Legendre_2020}. 
    The latter is especially important in time-dependent problems, where Green's functions depend on two time arguments and memory often becomes a limiting factor \cite{Balzer_Time_2010, Balzer_Efficient_2010, balzer_nonequilibrium_2012}. 
    In such settings, it is often useful to extend short-time simulations to longer times \cite{shi_new_2003,zhang_nonequilibrium_2006,cohen_memory_2011,cohen_numerically_2013,wilner_bistability_2013,cohen_generalized_2013,kelly_efficient_2013,wilner_nonequilibrium_2014,Cerrillo_Non_2014,wilner_sub-ohmic_2015,kidon_exact_2015,kelly_generalized_2016,rosenbach_efficient_2016,pollock_reduced_2022}.
    Compact representations are a particularly promising approach to this paradigm \cite{kaye_low_2021, yin_using_2022, Meirinhos_Adaptive_2022, blommel_adaptive_2024, sroda_high_2024}.
    
    While compact representations support extrapolation, predicting dynamics from limited-time data remains a central challenge. 
    A variety of approaches have been developed, including techniques that explicitly map past to future data, such as linear prediction \cite{makhoul_linear_1975, smith_extrapolation_1987, reynolds_investigation_1997, press_numerical_2007, Barthel_Spectral_2009, brezinski_extrapolation_2013} and dynamic mode decomposition (DMD) 
    \cite{Schmid_Dynamic_2010, yin_using_2022, Yin_Analyzing_2023, Reeves_Dynamic_2023, maliyov2024dynamic, kaneko_forecasting_2025}. 
    Other strategies forecast dynamics by minimizing entanglement entropy \cite{Lin_Novel_2024}, using machine learning models \cite{lin_simulation_2021, rodriguez_comparative_2022, Wang_Time_2024, zhu_predicting_2025}, or employing tensor networks \cite{Cerrillo_Non_2014, Gelzinis_Applicability_2017, Nunez_Learning_2022, Jeannin_Cross_2024}.
    When the goal is to extrapolate based on a compact representation, a distinct class of methods reconstructs the spectral content by expressing the time series as a sum of complex exponentials. 
    Prominent examples include Prony's method \cite{Prony_1795, Kumaresan_Prony_1984, Ribeiro_Non_2003, potts_parameters_2013}, the Matrix Pencil method \cite{hua_matrix_1990, sarkar_using_1995, Adve_Extrapolation_1997}, the Estimation of Signal Parameters via Rotational Invariance Techniques (ESPRIT) \cite{paulraj_subspace_1986, roy_esprit_1989, potts_parameters_2013, Stroeks_Spectral_2022, Shee_Real_2024, ding_esprit_2024}, and the Multiple Signal Classification (MUSIC) algorithm \cite{schmidt_multiple_1986, liao_music_2016}, while DMD can also be interpreted in this framework. 
    Each method comes with unique strengths and limitations, as explored in comparative studies \cite{Pogorelyuk_Clustering_2018}. 
    Continued refinement of these tools is key to advancing long-time predictions and deepening our understanding of quantum dynamics.
    
    In this paper, we investigate the applicability of the ESPRIT algorithm for representing real-time data from nonequilibrium quantum systems as a sum of complex exponentials. This compact form enables data-driven, physics-agnostic extrapolation of the system's dynamics. We evaluate ESPRIT's robustness to noise and benchmark its performance against common extrapolation methods, including linear prediction, DMD, and recurrent neural networks (RNNs). Additionally, we introduce a criterion based on the extracted exponentials to determine whether the short-time data contains sufficient information for reliable extrapolation. Finally, we discuss how this approach can be integrated into numerical propagation schemes and used to characterize long-time behavior in quantum systems.

    The outline of this paper is as follows: 
    In Sec.~\ref{sec:algorithms}, we introduce the exponential representation (Sec.~\ref{sec:compact_reps}), describe our ESPRIT algorithm (Secs.~\ref{sec:esprit} Sec.~\ref{sec:post_processing}), and review other common methods (Sec.~\ref{sec:other_alogs}). 
    Our results are presented in Sec.~\ref{sec:results}. 
    In Sec.~\ref{sec:noisy_test}, we showcase the susceptibility of all algorithms to noise using an analytical test function and analyze the performance of ESPRIT in the presence of noise in Sec.~\ref{sec:ESPRIT_noise_analysis}. 
    Sec.~\ref{sec:MC_extension} proposes an ESPRIT-based approach to reliably predict long-time dynamics from short-time data, including a criterion to assess if all relevant information is captured in the provided real-time data and how to integrate this into numerical propagation schemes. 
    In Sec.~\ref{sec:localization}, we demonstrate how ESPRIT can extract infinite-time limits from short-time data, enabling the study of localization behavior. 
    Sec.~\ref{sec:summary} provides a summary and outlook.

\section{Algorithms -- calculating compact representations in terms of complex exponentials}\label{sec:algorithms}

    In this section, we explain how to represent a function as a sum of exponential terms, introduce the ESPRIT algorithm for extracting these components from sampled data, and briefly review alternative methods for both exponential extraction and common extrapolation techniques.
    We refer to the resulting set of exponentials as the representation and focus on algorithms designed to reproduce the original signal as accurately as possible, while maintaining robustness and stability in the presence of noise. 
    We also outline strategies for filtering and postprocessing the extracted exponentials, which are particularly useful when working with noisy data.

    \subsection{Representation in terms of exponentials}\label{sec:compact_reps}

        The basic assumption of the methodology is that the dynamical observable of interest $f(t)$ is well approximated by a sum of $M$ exponentials,
        \begin{eqnarray}
            f(t) &=& \sum_{p=1}^{M} C_p e^{\xi_p t} .\label{eq:compact_exp}
        \end{eqnarray}
        Relating this form to quantum mechanical systems, the complex exponentials $\xi_p$ encode the system's characteristic frequencies -- or energies -- through their imaginary parts, and the life- or coherence times through their real parts. 
        The corresponding coefficients $C_p$ determine the amplitude or weight with which each exponential contributes to the overall signal. 
        This ansatz is well-justified for both equilibrium dynamics and nonequilibrium relaxation processes, as  $C_p$ and $\xi_p$ remain time-independent in both cases.

        Given the set of exponentials and prefactors $\{\xi_p, C_p\}$, the function $f(t)$ can be reconstructed at arbitrary times.
        In typical applications, these components are extracted from a finite set of discretized $f(t)$-values. 
        Since the number of exponentials is often significantly smaller than the number of time points in numerical or experimental datasets, we denote $\{\xi_p, C_p\}$ as the compact representation of the signal.
        Moreover, representing the data as a sum of complex exponentials has several advantages for postprocessing, as operations such as calculating Fourier transforms become straightforward, see App.~\ref{sec:app_FT} for details.

    \subsection{ESPRIT algorithm}\label{sec:esprit}

        The main focus of this work is the ESPRIT algorithm, a subspace-based signal processing technique for efficiently estimating the frequencies and amplitudes of superimposed exponential signals from noisy data~\cite{paulraj_subspace_1986, roy_esprit_1989, potts_parameters_2013}. 
        We assume that the data is sampled on an equidistant time grid 
        $t_i = \Delta t \cdot i$ with $i = 0, 1, 2, \dots, N$, 
        where each data point is given by 
        $f_i = f(t_i) + e_i$, and $e_i$ denotes the noise or error at grid point $i$. 
        The ESPRIT algorithm then estimates the exponents $\xi_p$, the prefactors $C_p$, and the number of exponentials $M$ as defined in Eq.~(\ref{eq:compact_exp}).

        The ESPRIT algorithm starts by rearranging the given data set into a
        Hankel matrix, 
        \begin{eqnarray}
                H &=&
                \left[
                \begin{array}{cccc}
                f_0 & f_1 & \cdots & f_L \\
                f_1 & f_2 & \cdots & f_{L+1} \\
                f_2 & f_3 & \cdots & f_{L+2} \\
                \vdots & \vdots &\ddots & \vdots \\
                f_{N-L} & f_{N-L+1} & \cdots & f_{N} \\
                \end{array}
                \right] , \label{eq:Hankel}
        \end{eqnarray} 
        where we have introduced the parameter $L$, which is typically chosen in the range between $N/3$ and $N/2$ to minimize variance \cite{sarkar_using_1995};
        in this work, we use $L=0.4 N$.
        We note that formulations starting from the Toeplitz matrix are also possible \cite{ding_esprit_2024}.
        Applying the singular value decomposition (SVD) to the Hankel matrix,
        \begin{equation}
            H = U \Sigma V^\dagger, \label{eq:Hankel_SVD}
        \end{equation}
        %
        %$s_0\geq s_1\geq \dots \geq s_L\geq0$ form the diagonal matrix $\Sigma$ and indicate the relative importance of each component in reconstructing the Hankel matrix.
        %The number of exponentials $M$ is then estimated as the smallest index for which $s_{M+1} < \epsilon$, where $\epsilon$ reflects the noise level. 
        we obtain the unitary matrices $U$ and $V$, whose columns represent the principal components of the signal and span the column and row spaces of $H$, while the singular values 
        $s_0 \ge s_1 \ge \dots \ge s_L \ge 0$ form the diagonal matrix $\Sigma$ and indicate the relative importance of each component in reconstructing the Hankel matrix.  
        The number of exponentials $M$ is then estimated by discarding singular values individually based on a relative cutoff, the singular values are first normalized by the largest one, $s_i / s_0$, and $M$ is chosen as the number of singular values satisfying $s_i / s_0 > \epsilon$, where $\epsilon$ is a prescribed threshold, often reflecting the noise level.
        If $\epsilon$ is unknown, it can be inferred from the distribution of singular values. 
        Truncating to the first $M$ singular values preserves only the significant data components, thereby minimizing the number of exponentials and effectively filtering out noise.

        The core idea of ESPRIT is the rotational invariance of signal subspaces constructed from time-shifted data. To exploit this, we define the matrices
        $U_0 = U[0:N-L-1, :]$ and
        $U_1 = U[1:N-L, :]$
        which are formed by removing the last and first row of $U$, respectively.
        Since the matrices $U_0$ and $U_1$ span the same signal subspace but correspond to adjacent time steps, they are related by a non-singular rotation $\Phi$ such that
        $U_0\Phi = U_1$,
        where $\Phi$ encodes the phase evolution of the exponential components. 
        In practice, $\Phi$ is computed as 
        \begin{eqnarray}
            \Phi &=& (U_0)^{+} U_1 , \label{eq:rotation_matrix}
        \end{eqnarray}
        where $(U_0)^{+}$ is the pseudo-inverse of $U_0$.
        The complex exponents $\xi_p$ are then extracted from the eigenvalues of $\Phi$ as
        \begin{eqnarray}
            \xi_p = \frac{\ln(\text{eig}(\Phi)_p)}{\Delta t} . \label{eq:exponents}
        \end{eqnarray}
        In the standard ESPRIT algorithm, the prefactors $C_p$ are then obtained by solving the Vandermonde system
        \begin{eqnarray}
                \left[
                \begin{array}{cccc}
                1 & 1 & \cdots & 1 \\
                e^{\xi_1\Delta t} & e^{\xi_2\Delta t} & \cdots & e^{\xi_M\Delta t} \\
                e^{\xi_1\Delta t 2} & e^{\xi_2\Delta t 2} & \cdots & e^{\xi_M\Delta t 2} \\
                \vdots & \vdots &\ddots & \vdots \\
                e^{\xi_1\Delta t N} & e^{\xi_2\Delta tN} & \cdots & e^{\xi_M\Delta tN}
                \end{array}
                \right] 
                \cdot
                \left[
                \begin{array}{c}
                    C_1\\
                    C_2\\
                    C_3\\
                    \vdots\\
                    C_M
                \end{array}
                \right]
                &=&
                \left[
                \begin{array}{c}
                    f_0\\
                    f_1\\
                    f_2\\
                    \vdots\\
                    f_N
                \end{array}
                \right]
                \label{eq:Vandermonde}
                ,
        \end{eqnarray}
        which reduces to a least squares problem when the number of data points exceeds the number of exponentials. In that case, the system approximates the signal in the form of Eq.~(\ref{eq:compact_exp}) by minimizing the residual error. A generalization of the ESPRIT algorithm to matrix-valued quantities can be found in Ref.~\cite{Ying_pole_2022_a, zhang_minimal_matrix_2024}. 
        Theoretically, it has been shown that ESPRIT is noise-resilient and can achieve an optimal error decay rate of $O(N^{-3/2})$~\cite{ding_esprit_2024} under certain assumptions on the noise structure. 
        Moreover, a recent application of the ESPRIT algorithm to quantum embedding and open quantum systems~\cite{park2024quasi} demonstrates that it yields the minimal number of poles necessary to achieve a target accuracy in hybridization function fitting. Similar findings have been reported in other recent studies~\cite{Takahashi_High_2024,ben-asher_memory_2024,lednev_lindblad_2024,zhang_minimal_2025}.
        We note that while ESPRIT identifies a set of discrete exponentials in a signal and therefore performs well when the signal itself contains only a few distinct exponentials, it can, in some cases, also represent signals composed of a continuum of exponentials surprisingly well, as briefly exemplified in App.~\ref{sec:app_FT}.

    \subsection{Postprocessing of ESPRIT exponents}\label{sec:post_processing}
        %Postprocessing ca be used to improve the robustness of the ESPRIT algorithm in the presence of noise and to tailor it to the physical problem at hand. Throughout the manuscript, we mainly emply two pPostprocessing steps, detailed in thefollowing. As trhe effectiveness of these thigs depend on teh qeustion and data at hand, we are selective when it comes to when to do what postprocessing, and we will specify when we use tyhem in teh correspondig sections in Sec.~\ref{sec:results}.
        %The firt postprocessing step whcih improves the robustness of the ESPRITR method for long times especially for noisy data is applied after computing the exponents in Eq.~(\ref{eq:exponents}) and before solving the Vandermonde system in Eq.~(\ref{eq:Vandermonde}). 
        %Specifically, we discard all exponents with a positive real part, corresponding to exponentially growing contributions. 
        Postprocessing can be used to enhance the robustness of the ESPRIT algorithm in the presence of noise and to tailor it to the specific physical problem. In this manuscript, we mainly employ two postprocessing steps, described below. Their application depends on the data and the question at hand, and we indicate throughout Sec.~\ref{sec:results} when each step is applied.  

        The first postprocessing step, which improves the robustness of ESPRIT for long-time extrapolation, particularly in noisy conditions, is applied after computing the exponents in Eq.~(\ref{eq:exponents}) and before solving the Vandermonde system in Eq.~(\ref{eq:Vandermonde}). Specifically, all exponents with a positive real part, which correspond to exponentially growing contributions, are discarded.
        While such terms may improve short-time fits, they lead to unphysical behavior in long-time extrapolations. 
        This filtering is especially important when extrapolating from noisy data. Additional filtering strategies -- such as rejecting high-frequency exponents or imposing other physically motivated constraints -- 
        can further enhance performance. 
        Although we restrict our filtering to removing exponentially increasing components in this work (unless we state otherwise), we emphasize that incorporating domain-specific knowledge to refine the exponent set can significantly improve the accuracy and reliability of ESPRIT, albeit at the cost of generality.
        Similar ideas of restricting solutions to physically meaningful ones have been explored for other methods \cite{Baddoo_Physics_2023}, and applying such principles to ESPRIT could similarly enhance predictive accuracy.

        The second postprocessing step addresses situations where access to the long-time behavior of a system is important, as is often the case in physics applications when characterizing quantities such as steady-state populations, residual polarizations or magnetizations, and persistent currents.
        In this work, the infinite-time limit $f(t\rightarrow\infty)$ serves as a key figure of merit to assess extrapolation accuracy (see Sec.~\ref{sec:noisy_test}) and to study localization phenomena (see Sec.~\ref{sec:localization}).
        To reliably estimate this asymptotic value, we explore two strategies:
        (i) adding an explicit zero exponent before solving the Vandermonde system, or
        (ii) setting the exponent with the smallest absolute value to zero prior to the fit.
        In both cases, the infinite-time value is given by the prefactor corresponding to the zero exponent. 
        While both approaches perform comparably overall, method (i) requires longer sampling times in high-noise regimes, whereas method (ii) is faster but slightly more susceptible to distortion of long-time predictions.
        We observed that these strategies are most susceptible to predicting different behaviors when slowly decaying exponents of small magnitude are present, particularly if their associated prefactors are comparable to the noise level. 
        In such cases, setting the smallest exponent to zero can remove information encoded in the data, typically giving accurate short-time behavior but inaccurate long-time predictions, whereas adding a zero exponent increases the effective complexity of the fit and generally requires longer time series to produce reliable results.
        Given that general statements depend on the available data, monitoring the predictions as a function of available data, as done in Secs.~\ref{sec:MC_extension} and \ref{sec:localization}, allows one to assess which approach is more appropriate for a given situation.

        We note that ESPRIT is a data-driven method, and its predictive power depends on the available data and the specific question at hand, as   illustrated by our results in Sec.~\ref{sec:results}. 
        While our focus here is on long-time behavior and steady states, other phenomena that can arise on long timescales include metastable states and multi-stability, which appear in various physical contexts. 
        In principle, ESPRIT should identify metastable behavior if it is encoded in the short-time data.
        We explore this to some extent in Sec.~\ref{sec:localization}, where we assess whether the data predicts a long-time decay to zero or convergence to a finite value. 
        The reliability of such predictions, and the amount of data required, depends strongly on the system studied, making this a problem-specific question beyond the scope of the present work. 
        A detailed investigation of ESPRIT's ability to identify metastable and multistable behavior under different conditions would be an interesting direction for future research.

        In the remainder of this work, we indicate which postprocessing steps are used in each case. 
        We also note that other strategies for extracting the long-time value may be viable, especially when additional information about the system or the noise is available. 
        Moreover, these postprocessing steps are not limited to the ESPRIT algorithm and can be integrated into any method that extracts exponentials from data and fits them accordingly.

    \subsection{Overview of other algorithms for extrapolating real-time data}\label{sec:other_alogs}

        In this work, a key application is the extrapolation and long-time prediction from short-time data. 
        Beyond ESPRIT, many alternative methods -- such as linear prediction \cite{makhoul_linear_1975, smith_extrapolation_1987, Barthel_Spectral_2009}, DMD \cite{Schmid_Dynamic_2010, yin_using_2022, Yin_Analyzing_2023, Reeves_Dynamic_2023, maliyov2024dynamic, kaneko_forecasting_2025}, and RNNs \cite{zhu_predicting_2025,bassi2024learning} -- have also been used for data extrapolation. 
        While many of these data-driven approaches share closely related theoretical foundations, they differ in how the signal exponents and coefficients are numerically estimated, leading to variations in predictive performance~\cite{Pogorelyuk_Clustering_2018}. 
        To place our method in context, we briefly introduce these alternatives and compare their predictive performance in Sec.~\ref{sec:noisy_test}. We note that each method has various generalizations; however, for both the introductory overview and the comparative study, we focus on their \textit{most basic formulations}.
        Implementing more advanced versions of these methods and conducting a detailed comparison to optimize their performance are beyond the scope of this work.
        \\

        {\em (a) Linear prediction --} One of the most widely used extrapolation methods is the \textit{linear prediction framework} \cite{makhoul_linear_1975, smith_extrapolation_1987, Barthel_Spectral_2009}, which estimates future values based on prior observations. 
        It models a given data set $f_i$ as a linear combination of its past values $f_{i-k}$,  
        \begin{equation}
            f_i = \sum_{k=1}^p a_k f_{i-k},
        \end{equation}
        where $p$ is the prediction order and $\{a_k\}$ are the prediction coefficients. 
        These coefficients are typically obtained by minimizing the mean squared prediction error,
        $\epsilon = \sum_{n=p+1}^{N} \left| f_n - \sum_{k=1}^p a_k f_{n-k} \right|^2$,
        which leads to the Yule-Walker equations,
        $\sum_{k=1}^p a_k R(j-k) = -R(j), \quad j=1,\dots,p$,
        with the autocorrelation function defined as
        $R(j) = \sum_{n=j+1}^{N} f_n f_{n-j}$.
        Linear prediction performs well for analytic and periodic functions \cite{Hasselmann_Techniques_1981}, but its accuracy degrades for non-analytic functions \cite{bree_prediction_2013} and signals with long-range correlations or nonstationary behavior \cite{godet_linear_2009}. 
        It is also sensitive to noise \cite{Koehl_Linear_1999, bree_prediction_2013,lee_robust_1988, Lin_Novel_2024} leading to under- or overfitting. \\

        {\em (b) Dynamic Mode Decomposition --} DMD is a data-driven method for learning and predicting dynamics, which has recently gained popularity in the condensed matter physics community for analyzing both equilibrium and nonequilibrium quantum dynamics~\cite{Schmid_Dynamic_2010, yin_using_2022, Yin_Analyzing_2023, Reeves_Dynamic_2023, maliyov2024dynamic, kaneko_forecasting_2025}. 
        Unlike signal processing approaches, DMD seeks to approximate the effective one-time propagator \( \mathcal{K}(\Delta t) \), where \( \mathcal{K} \) denotes the Koopman operator governing the evolution of the observable \( \mathbf{x}(t) \). 
        To this end, one collects data snapshots
        \begin{eqnarray}
            X = \begin{bmatrix} \mathbf{x}_0 & \mathbf{x}_1 & \cdots & \mathbf{x}_{m-1} \end{bmatrix}, \
            X' = \begin{bmatrix} \mathbf{x}_1 & \mathbf{x}_2 & \cdots & \mathbf{x}_m \end{bmatrix},
        \end{eqnarray}
        and determines a best-fit linear operator \( A \approx \mathcal{K}(\Delta t) \) satisfying \( X' \approx A X \). 
        The least-squares solution is given by \( A = X' X^+ \). To reduce dimensionality, one performs a rank-\( r \) SVD
        $X \approx U_r \Sigma_r V_r^\dagger$,
        and defines the reduced operator,
        \begin{eqnarray}
            \tilde{A} = U_r^\dagger X' V_r \Sigma_r^{-1}  \label{eq:reduced_op}.
        \end{eqnarray}
        The matrix \( \tilde{A} \) is the projection of \( A \) onto the subspace spanned by the leading \( r \) proper orthogonal decomposition modes, and provides a finite-dimensional approximation of \( \mathcal{K}(\Delta t) \). 
        The dynamics of an observable \( \mathbf{x}(t) \) can then be approximated via the eigendecomposition \( \tilde{A} = W \Lambda W^{-1} \) as:
        \begin{eqnarray}
            \mathbf{x}(t) \approx \tilde{\mathbf{x}}(t) = \sum_{j=1}^r b_j \phi_j e^{\omega_j t}, \ \text{where } \omega_j = \frac{\log(\lambda_j)}{\Delta t}.
        \end{eqnarray}
        Here, the DMD modes in the original space are computed as \( \Phi = X' V_r \Sigma_r^{-1} W \), with each column \( \phi_j \in \mathbb{C}^n \) corresponding to the eigenvalue \( \lambda_j \) of \(\tilde{A}\). 
        The coefficients \( \mathbf{b} = \Phi^+ \mathbf{x}_0 \) represent the projection of the initial condition onto the DMD modes. 
        
        In our application, we adopt a variant known as \emph{high-order DMD} (HO-DMD) ~\cite{le2017higher}. 
        In this approach, the observable \( f(t) \) is delay-embedded by stacking \(n_s \) successive values to construct a vector
        $\mathbf{x}_0 = [f_0; f_1; \cdots; f_{n_s-1}]$,
        after which the standard DMD procedure is applied to the resulting time-delay embedded snapshots. 
        HO-DMD is particularly useful for low-dimensional systems—such as the scalar case where \( \mathbf{x}(t) = f(t) \in \mathbb{C} \)—where the rank of the matrix \( A \) is otherwise too low to extract meaningful dynamical information. 
        Compared to ESPRIT, the DMD method is generally more sensitive to noise~\cite{Pogorelyuk_Clustering_2018, Baddoo_Physics_2023}, although generalizations such as the noise-resistant DMD have been proposed to mitigate this issue ~\cite{wanner2022robust}.
        Since the ESPRIT approach is formulated based on the Hankel matrix, it is most directly comparable to the Hankel-DMD method~\cite{arbabi2017ergodic}. 
        A detailed comparative analysis between the two approaches is provided in App.~\ref{sec:append_DMD_ESPRIT_comnparison}.

        {\em (c) Recurrent neural network --} Machine learning methods can also be employed to learn the effective time propagator governing the dynamics of an observable \( f(t) \). 
        To this end, various neural network architectures may be used. 
        In this work, we adopt a recurrent neural network (RNN) approach similar to the formulations in Refs.~\cite{zhu_predicting_2025,bassi2024learning}. 
        Specifically, we use an Long Short-Term Memory (LSTM) based RNN to model the time derivative \( f'(t) \) as
        \begin{align}
            f(t) \xrightarrow{\text{RNN}} f'(t) .
        \end{align}
        Once the mapping is learned from short-time trajectories of \( f(t) \), the future evolution of the system can be obtained using a numerical integration scheme. For example, applying the Euler method yields
        \begin{align}
            f(T+\Delta t) = f(T) + \Delta t \cdot \text{RNN}(f(T)) .
        \end{align}
        RNN-based ML method has shown numerical advantages for learning/predicting high-dimensional dynamical systems \cite{zhu_predicting_2025}.\\

        {\em (d) Other subspace/signal processing methods --} Besides ESPRIT, there are several other methods to estimate complex exponentials from data, including Prony's approximation method \cite{Prony_1795, Beylkin_approximation_2005, Beylkin_approximation_2010}, the Matrix Pencil method \cite{sarkar_using_1995, hua_matrix_1990, potts_parameters_2013}, and subspace approaches such as Multiple Signal Classification (MUSIC) \cite{schmidt_multiple_1986, liao_music_2016}. 
        Prony’s method is simple and direct; although its approximation variant is generally stable \cite{Beylkin_approximation_2005, Beylkin_approximation_2010}, interpolation using Prony’s method is highly sensitive to noise \cite{moitra_super_2015}. In contrast, the Matrix Pencil method offers improved robustness to noise but can struggle with closely spaced or heavily damped modes \cite{hua_matrix_1990}.
        Subspace methods like ESPRIT and MUSIC are generally more noise-resilient and numerically stable, as they rely on subspace projections rather than root-finding or solving generalized eigenvalue problems. 
        In contrast to other subspace methods, ESPRIT constructs a Hankel matrix and exploits the rotational invariance of the signal subspace to extract the exponents directly, whereas other methods such as MUSIC use the correlation matrix and identify frequencies via a spectral peak search \cite{roy_esprit_1989}. 
        This makes ESPRIT particularly efficient when both damping and phase information are needed, as it avoids exhaustive searches and yields full complex exponentials directly. 
        We focus exclusively on ESPRIT for exponential estimation in this work.

\section{Results -- Extrapolating real-time data}\label{sec:results}

    We begin by benchmarking ESPRIT using an analytic test function. 
    Sec.~\ref{sec:noisy_test} compares ESPRIT to other standard extrapolation techniques, and Sec.~\ref{sec:ESPRIT_noise_analysis} analyzes its robustness and performance in the presence of noise.
    Subsequently, we apply ESPRIT to Quantum Monte Carlo (QMC) data. 
    Sec.~\ref{sec:MC_extension} shows how restricted propagators for the Anderson impurity model in the correlated regime can be extended to longer times, based on a criterion that determines when all information has been extracted from the short-time dynamics.
    Sec.~\ref{sec:localization} focuses on spin-polarization in the spin-boson model, demonstrating that long-time localization effects can be inferred from short-time data, and assessing whether the available data is sufficient for a reliable prediction.

    \subsection{Extrapolating noiseless and noisy data} \label{sec:noisy_test}
        
        \begin{figure*}[tb]
            \raggedright (a)\\
            \centering
            \includegraphics{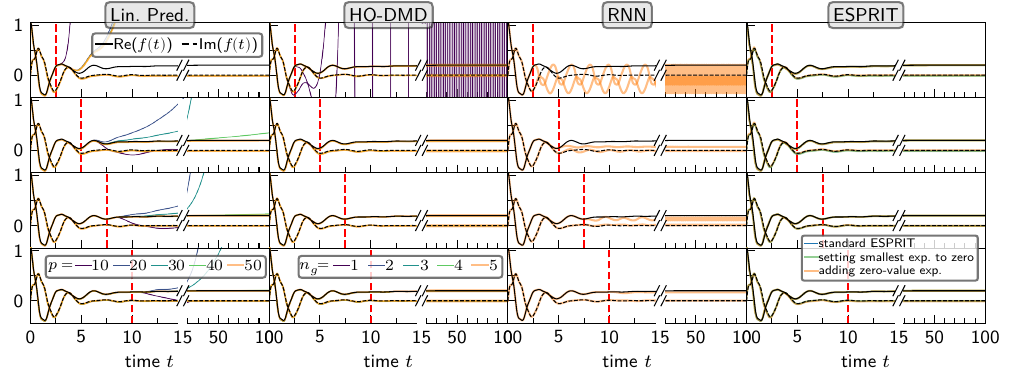}
            \\ \raggedright (b)\\
            \centering
            \includegraphics{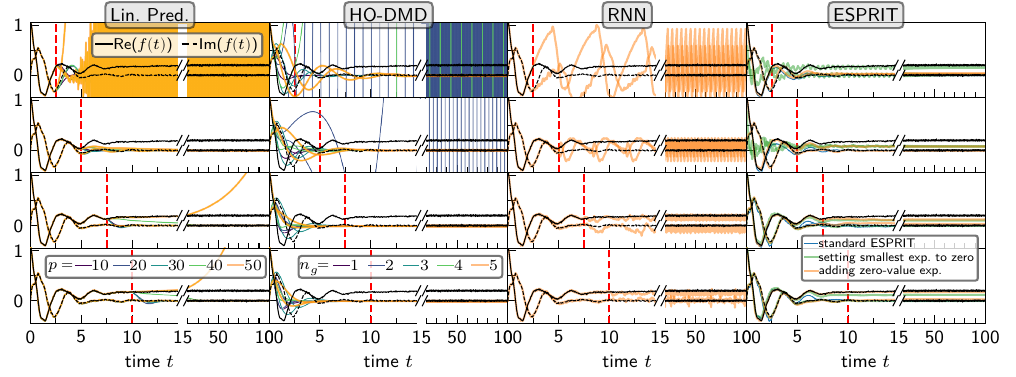}
            \caption{
                Extrapolation of the function $f(t)$ (Eq.~(\ref{eq:test_func})) using different methods (from left to right: linear prediction, higher-order dynamic mode decomposition (HO-DMD), recurrent neural network (RNN), and ESPRIT).
                The function $f(t)$ is shown as solid and dashed black lines for the real and imaginary parts, respectively. Colored lines represent predictions from various methods: different orders $p$ for linear prediction and $n_g$ for HO-DMD, as well as different strategies for incorporating the final value in ESPRIT. 
                The threshold used for the SVD decomposition in both HO-DMD and ESPRIT is $10^{-6}$.
                The details of the RNN used to predict the dynamics are specified in App.~\ref{sec:append_ML_dyn}.
                Shaded regions correspond to extrapolations with rapid oscillations, causing the area to appear filled.
                Panels (a): Without noise.
                For ESPRIT and HO-DMD, the predictions are so accurate that they coincide with the black lines representing the analytical test function $f(t)$.
                Panels (b): With Gaussian noise with $\sigma=10^{-2}$. 
                In both panels, the sampling time $t_{\text{samp}}$ (red vertical dashed line), that is the time up to which the data is available to the algorithms, increases from top to bottom
                with $t_{\text{samp}}=2.5,\ 5.0,\ 7.5,$ and $10.0$, respectively.
            }
            \label{fig:method_comparison}
        \end{figure*}

        We begin our analysis by comparing the extrapolation performance of ESPRIT with other common methods introduced in Sec.~\ref{sec:other_alogs}, using the analytic test function
        \begin{eqnarray}
            f(t) &=& c_1 e^{-o_1 t} + c_2 e^{-o_2 t} + c_3 e^{-o_3 t} + c_4 e^{-o_4 t} + f_\infty , \nonumber \\
            \label{eq:test_func}
        \end{eqnarray}
        with parameters $c_1 = 0.85$, $c_2=0.15$, $c_3=0.1$, $c_4=-0.2$ and $o_1=1/2-2i$, $o_2=1/3+3i$, $o_3=1-10i$, $o_4=0.2$, and $f_\infty=0.2$ for our test-case. 
        This analytic function approaches a finite asymptotic value, $f(t \rightarrow \infty) = f_\infty = 0.2$.
        We evaluate $f(t)$ on an equidistant time grid with step size $\Delta t = 0.025$. To investigate robustness under noise, we add complex Gaussian noise with zero mean and standard deviation $\sigma$, equally distributed between the real and imaginary parts.

        Fig.~\ref{fig:method_comparison} showcases the extrapolation of the function $f(t)$, defined in Eq.~(\ref{eq:test_func}), using linear prediction, HO-DMD, a RNN, and ESPRIT. 
        Panels (a) show the noise-free case, while panels (b) depict results in the presence of Gaussian noise of standard deviation $\sigma = 10^{-2}$. 
        In all panels, the red dashed line marks $t_{\text{samp}}$, the cutoff time up to which data is provided to the algorithms,
        i.e., the region to the left of the red dashed line corresponds to known input data, while the region to the right shows the extrapolation. 
        From top to bottom, the amount of available data increases, which should enable a more reliable extrapolation.
        We note that Fig.~1 illustrates the general behavior of the different extrapolation methods for context, rather than providing a detailed analysis of their differences or applicability. 
        All methods are implemented in their basic form, except for ESPRIT, which discards exponentially growing contributions and also considers results from two nonstandard variants that account for finite-size effects, as discussed in Sec.~\ref{sec:post_processing}. 
        Extensions to improve performance exist for all methods but are beyond the present scope.

        In the noiseless case in Fig.~\ref{fig:method_comparison}(a), all methods extrapolate the correct behavior given sufficiently large $t_{\text{samp}}$. 
        Linear prediction tends to decay to zero or become unstable if $t_{\text{samp}}$ is short or the memory depth $p$ is too small. 
        Larger $p$ improves performance in such cases. 
        HO-DMD of order $n_s = 5$ is reliable except for the smallest $t_{\text{samp}}$ and $n_g$ ($n_g\Delta t$ is the sampling gap of the signal), where it becomes unstable since the learned frequency  $\omega_j=\log(\lambda_j)/(n_g\Delta t)$ becomes large for small $n_g$. 
        We note that increasing $n_s$ can substantially improve reliability and performance, and that applying DMD to the same input data as ESPRIT brings its performance much closer to that of ESPRIT, as demonstrated in App.~\ref{sec:append_data_HankelDMD}, where $n_s$ ranges from 60 to 240 depending on $t_{\text{samp}}$.
        The RNN captures the correct long-time behavior for large $t_{\text{samp}}$, but exhibits persistent oscillations or underestimates $f_\infty$ when too little data is provided. 
        ESPRIT is the most robust, meaning that it is the only method that consistently recovers the correct dynamics across all $t_{\text{samp}}$ for every parameter and data set considered, regardless of how the final value is incorporated.
        It also correctly identifies the number of exponentials, using a threshold of $\epsilon = 10^{-6}$ to filter singular values after the SVD. (The consequences of using an incorrect number of exponentials $M$ are analyzed in Fig.~\ref{fig:noise_analysis}).
        
        In the presence of noise, as shown in Fig.~\ref{fig:method_comparison}(b), 
        and using the same numerical parameters for all methods as in the noiseless case, 
        the extrapolation quality changes significantly. 
        Linear prediction becomes unstable for short $t_{\text{samp}}$ and, even with longer sampling times and larger $p$, fails to recover the correct infinite-time value $f_\infty$, typically predicting zero or diverging. 
        For this specific data set, the basic HO-DMD performs poorly throughout; however, when $n_s$ is substantially increased and the input data is organized in the same way as for ESPRIT, HO-DMD performs much better, even in the presence of noise (see Fig.~\ref{fig:DMD_ESPRIT_Fig1}(b) and the accompanying discussion).
        For short $t_{\text{samp}}$ and small $n_g$, it shows diverging oscillations, while for larger values, it predicts a decay to zero—even while failing to reproduce the known input dynamics. 
        We note that DMD performance can depend sensitively on the SVD truncation threshold, and that information-theoretic criteria have been proposed which may perform well under various noise distributions \cite{gavish_optimal_2014}.
        All numerical parameters common to HO-DMD and ESPRIT are identical, with the SVD threshold in HO-DMD also being set to $10^{-6}$.
        The RNN performs better with longer $t_{\text{samp}}$, but in the presence of noise, it amplifies the oscillatory behavior and underestimates $f_\infty$, as already seen in the noiseless case.
        ESPRIT remains the most robust method in the above sense, which also works as a denoising technique, but now its accuracy in recovering $f_\infty$ depends on how the zero exponent is incorporated. 
        Using the standard ESPRIT algorithm without adjustment, it tends to predict a decay to zero at long times. 
        Setting the smallest-value exponent to zero can improve long-time predictions but may compromise short-time accuracy, as the exponent set to zero may not correspond to the constant term $f_\infty$. 
        For our test case, including a zero exponent explicitly before fitting yields the most reliable results across time scales.
        
        We note that adding fully uncorrelated Gaussian noise as done in this test case is somewhat artificial, as real-world noise typically exhibits temporal correlations. This represents a worst-case scenario for methods like linear prediction and DMD, which rely on a connection between past and future time steps. 
        While more robust algorithmic variants exist~\cite{wanner2022robust}, we restrict our study to the standard algorithm.
        For DMD practitioners interested in a more detailed discussion of the differences, we provide a direct comparison between Hankel-DMD and ESPRIT in App.~\ref{sec:append_data_HankelDMD}.

    \subsection{Performance of the ESPRIT algorithm in the presence of noise}\label{sec:ESPRIT_noise_analysis}
    
        \begin{figure}[tb]
            \raggedright (a) \hspace{0.45\linewidth}(b)\\
            \centering            \includegraphics{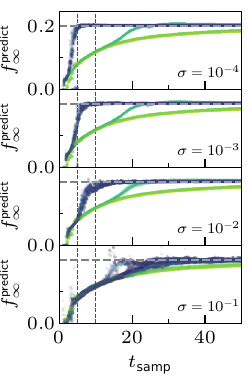}            \includegraphics{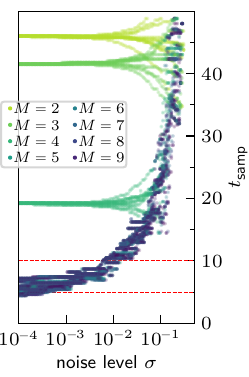}\\
            \raggedright (c)\\ \centering
            \hspace{-1cm}\includegraphics{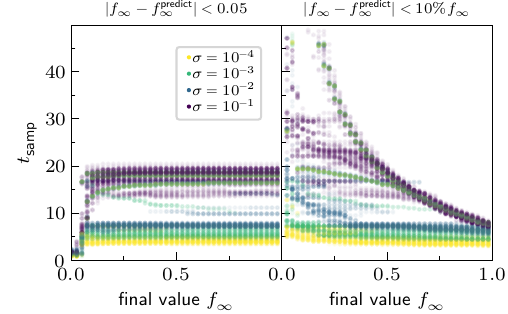}
            \caption{
                Influence of noise on the fidelity of long-term predictions by ESPRIT.
                Panels (a): Predicted values of $f_{\infty}^{\text{predict}}$ as a function of sampling time $t_{\text{samp}}$ for different noise strengths $\sigma$ and numbers of exponents $M$ (including the zero exponent).
                The gray dashed line indicates the true value of $f_\infty=0.2$.
                Panel (b): Sampling time $t_{\text{samp}}$ required to achieve an accuracy of $|f_\infty-f_\infty^{\text{predict}}|<0.02$, shown for various values of $M$ (which includes the zero exponent).
                The red dashed lines in panels (a) and (b) highlight the values $t_{\text{samp}}=5.0$ and $t_{\text{samp}}=10.0$ and serve as a guide for the eye and indicate a regime of $t_{\text{samp}}$ for which an accurate extrapolation is desirable.
                Panels (c): Influence of the final value $f_\infty$ on the predictability of the ESPRIT algorithm at different noise levels.   
                Sampling time required to reach a fixed accuracy of $0.05$ (left) and a relative accuracy of $10\%$ of $f_\infty$ (right), whereby colors indicate different noise levels $\sigma$.
                All three panels show results from ten independent runs with distinct random noise seeds.
            }
            \label{fig:noise_analysis}
        \end{figure}

        Having demonstrated ESPRIT's robust performance in the presence of noise, we now provide a more systematic analysis of how noise strength affects the sampling time $t_{\text{samp}}$ required for accurate predictions. 
        %We also investigate the role of the number of exponents $M$ used within the ESPRIT algorithm, rather than setting tolerance as we did in the previous section. 
        %To this end, we explicitly include a zero exponent and as before, discard exponentially diverging contributions, as discussed in Sec.~\ref{sec:post_processing}.
        We also investigate the role of the number of exponents $M$ used within the ESPRIT algorithm, rather than determining it via a tolerance as in the previous section. To this end, we explicitly include a zero exponent and, as before, discard exponentially diverging contributions, as described in Sec.~\ref{sec:post_processing}.
        As a figure of merit, we focus on the predicted value $f_\infty^{\text{predict}}$ and compare it to the true value $f_\infty$, which is both straightforward to compute and related to physical observables (see Sec.~\ref{sec:localization}). 
        As before, we use the analytic test function defined in Eq.~(\ref{eq:test_func}) and add Gaussian noise of standard deviation $\sigma$.

        Fig.~\ref{fig:noise_analysis} studies how the required sampling time for reliable predictions of $f_\infty$ depends on the noise strength $\sigma$ and the number of exponents $M$ used in ESPRIT, with $M$ including the zero exponent.
        Fig.~\ref{fig:noise_analysis}(a) plots the predicted $f_\infty^{\text{predict}}$ versus $t_{\text{samp}}$, with different colors denoting different values of $M$. Red dashed lines serve as visual guides: a reliable prediction at $t_{\text{samp}} < 5$ is highly valuable, as the function has not yet visibly plateaued, while predictions beyond $t_{\text{samp}} > 10$ are less useful, since the function is already close to its final value.
        In the top three panels (low to intermediate noise), ESPRIT accurately recovers $f_\infty$ for $t_{\text{samp}} \gtrsim 5$ if $M \geq 5$, with only a mild increase in required $t_{\text{samp}}$ as noise increases. 
        For $M < 5$, a much longer sampling time is needed because the test function contains five exponential components, and not allowing for sufficient components must be compensated for by relying on long-time features that dominate late dynamics.
        At high noise levels (bottom panel), ESPRIT fails to predict $f_\infty$ accurately unless much longer $t_{\text{samp}}$ is provided. Increasing $M$ no longer helps, as separating signal from noise becomes challenging. 
        In this regime, ESPRIT functions more as a noise filter, producing reasonable results only when $f(t)$ has effectively reached its long-time value and the main challenge is denoising rather than extrapolation.
        Relating these findings to the previous section, we note that while the RNN was not the most stable method in the previous section, it is in principle possible to train a neural network specifically to predict only the final value, enabling a machine learning variant of the analysis presented here. For completeness, we include a brief discussion of this approach in App.~\ref{sec:append_ML_inf}.

        Fig.~\ref{fig:noise_analysis}(b) quantifies how the sampling time $t_{\text{samp}}$ required to recover $f_\infty$ depends on the noise strength $\sigma$. 
        The panel shows the minimum $t_{\text{samp}}$ needed to predict $f_\infty$ within $0.02$ of its true value, that is 
        the difference between $f_\infty$ and $f_\infty^{\text{predict}}$ is smaller than $0.02$. 
        For $M \geq 5$ and $\sigma \lesssim 2\cdot10^{-2}$, ESPRIT yields accurate results with only a mild increase in $t_{\text{samp}}$ as noise grows. 
        However, around $\sigma \sim 10^{-1}$, the required sampling time diverges, indicating that noise begins to dominate the data.
        In this regime, ESPRIT can no longer reliably identify the true signal and primarily detects noise, which it attempts to filter out rather than extract the meaningful exponents needed for extrapolation.
        For $M < 5$, the required $t_{\text{samp}}$ remains roughly constant at a high value even for low noise, reflecting ESPRIT's limited model complexity due to the insufficient number of exponentials.
        As $\sigma$ increases, $t_{\text{samp}}$ eventually also diverges as noise starts to dominate the data.

        In practical applications, the signal-to-noise ratio, or how much noise the approach can tolerate relative to the value of the observable $f_\infty$, can be important.
        This question is analyzed in Fig.~\ref{fig:noise_analysis}(c), which shows the required sampling time to reach either a fixed absolute precision of $0.05$ (left panel) or a relative precision of $10\%$ of the final value $f_\infty$ (right panel). 
        Unlike previous figures where $f_\infty = 0.2$, here $f_\infty$ ranges from $0$ to $1$, evaluated at four representative noise levels.
        For the fixed precision (left panel), $f_\infty$ has little effect on the required $t_{\text{samp}}$, which depends almost entirely on the noise level. 
        This reflects that ESPRIT first estimates exponents rather than prefactors, making it less sensitive to the absolute value of $f_\infty$. 
        An exception to this is very small $f_\infty \leq 0.05$, where zero falls within the allowed tolerance, and ESPRIT tends to default to predicting $f_\infty^{\text{predict}} = 0$ when $t_{\text{samp}}$ is too short (see Fig.~\ref{fig:noise_analysis}(a)).
        In contrast, when accuracy is defined as a percentage of $f_\infty$ (right panel), the required $t_{\text{samp}}$ decreases with increasing $f_\infty$, simply because the tolerance grows with the signal's magnitude.

    \subsection{Prediction of long-time dynamics from short-time data in numerical propagation scheme}\label{sec:MC_extension}

        Beyond predicting infinite-time values, ESPRIT's ability to extrapolate noisy data is valuable for both experimental analysis and theoretical studies of time-dependent processes. 
        This is particularly relevant for strongly correlated systems, where coherence times can exceed intrinsic dynamical timescales by orders of magnitude. 
        Simulating such systems is challenging, as accessing long-time behavior scales at least linearly with the coherence time, and time-propagation or resummation schemes often accumulate errors over time.
        In this context, reliably inferring long-time dynamics from short-time data is highly desirable, and ESPRIT offers a systematic approach. 
        We therefore develop a strategy to identify when the short-time signal contains all relevant information, such that extrapolation becomes more efficient than continued time-propagation.
        To investigate this, we move beyond analytic test cases and apply ESPRIT to real-time data obtained from continuous-time QMC simulations \cite{Gull_Continuous_2011} of the single-orbital Anderson impurity model at various temperatures, probing both the uncorrelated and strongly correlated regimes.

        We consider the single-orbital Anderson impurity model described by the Hamiltonian $H = H_\text{I} + H_\text{B} + H_\text{IB}$, with the interacting impurity 
        $H_\text{I} =  \sum_{\sigma} \epsilon_0 d_\sigma^\dagger d_\sigma + U d_\uparrow^\dagger d_\uparrow d_\downarrow^\dagger d_\downarrow$, the noninteracting bath 
        $H_\text{B} = \sum_{\sigma k} \epsilon_{k} a_{k\sigma}^\dagger a_{k\sigma}$, 
        and the coupling between impurity and bath 
        $H_\text{IB} = \sum_{\sigma k} \left( V_{k} a_{k\sigma}^\dagger d_\sigma + \text{h.c.} \right)$, 
        which is also referred to as hybridization.
        Here, $\epsilon_0$ is the on-site energy on the impurity and $U$ is the Coulomb interaction strength.
        $d^{(\dagger)}_\sigma$ are fermionic creation and annihilation operators with spin $\sigma\in\lbrace \uparrow, \downarrow  \rbrace$ for the impurity, and $a^{(\dagger)}_{k\sigma}$ are their counterparts on the bath orbitals with energy $\epsilon_k$.
        The coupling between the impurity and the bath is often parametrized by the coupling strength function 
        $\Gamma(\epsilon) = 2\pi \sum_{k} |V_{k}|^2 \delta(\epsilon-\epsilon_{k})$,
        which can model the specifics of a nanoscale system  \cite{Erpenbeck_Shaping_2023} or an effective DMFT bath for a lattice problem \cite{Georges_Dynamical_1996, Vollhardt_Dynamical_2012, Vollhardt_Dynamical_2012_2, Aoki_Nonequilibrium_2014}.

        \begin{figure}[h!]
            \centering
            \raggedright 
            (a) \hspace{1.7cm}\textbf{without filtering of exponents}\\
            \includegraphics{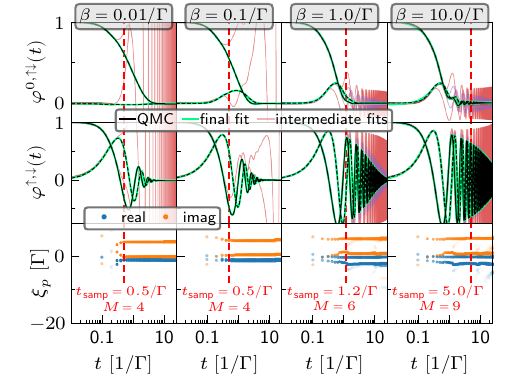}\\
            \vspace{0.33cm}
            \raggedright 
            (b) \hspace{0.2cm}\textbf{excluding exponentially increasing contributions}\\\includegraphics{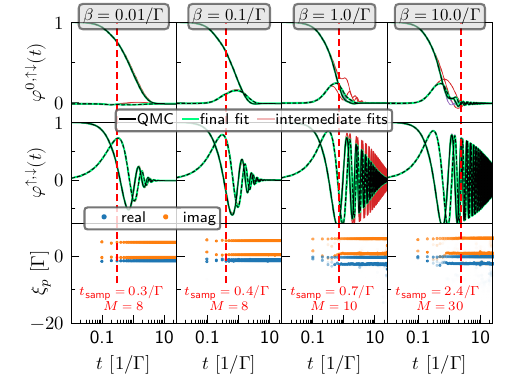}
            \caption{
                Application of the ESPRIT algorithm to QMC data across increasing inverse temperatures $\beta$ (left to right).
                Panels (a) show results for the standard ESPRIT algorithm with all exponents included and no additional postprocessing, while panels (b) show results where exponentially increasing components have been removed (see Sec.~\ref{sec:post_processing} for details).
                In both panels, the top two rows show the particle-hole symmetric restricted propagators $\varphi$, with black solid and dashed lines representing the real and imaginary parts obtained from QMC, respectively. 
                The red dashed vertical line marks the sampling time $t_{\text{samp}}$, where ESPRIT achieves a prediction accuracy of $10^{-3}$ with respect to the QMC data. 
                Bright green lines show ESPRIT extrapolations from this $t_{\text{samp}}$.
                Faint red lines indicate ESPRIT predictions using shorter sampling times $0.2/\Gamma$, $0.5/\Gamma$, $1.0/\Gamma$, and $2.0/\Gamma$, where applicable.
                The bottom row in both panels displays the extracted exponents $\xi_p$, which are shared by all restricted propagators, as a function of time provided to the algorithm, with scatter point opacity determined by the corresponding maximum absolute value of the prefactor $C_p$, making more significant exponents visually more prominent.
                The tolerance for the SVD decomposition in the ESPRIT algorithm was set to \(10^{-9}\).
            }
            \label{fig:propagators}
        \end{figure}
        
        For the scope of this work, we apply our extrapolation frameworks to restricted real-time propagators
        $\varphi^\alpha(t) = \text{Tr}_\text{B} \left\lbrace \rho_\text{B} \bra{\alpha}e^{-iHt}\ket{\alpha} \right\rbrace$,
        which we consider one of the most basic dynamical objects for a quantum system, and
        which are a central object for the real-time \cite{Cohen_Taming_2015, Antipov_Currents_2017, Chen_Inchworm_2017, Chen_Inchworm_2017_2, Cai_Inchworm_2020, cai_numerical_2023} and the nonequilibrium steady-state inchworm methodologies \cite{erpenbeck_quantum_2023, erpenbeck_steady_2024}.
        Here, $\text{B}$ denotes the bath degree of freedom and $\alpha\in\lbrace 0, \uparrow, \downarrow, \uparrow\downarrow\rbrace$ are the states of the single-orbital impurity.
        A detailed discussion of these propagators is given in Ref.~\cite{Cohen_Greens_2014, Erpenbeck_Resolving_2021}.
        We evaluate the restricted propagators using the real-time inchworm QMC method \cite{Cohen_Taming_2015,Antipov_Currents_2017,dong_quantum_2017,boag_inclusion-exclusion_2018,ridley_numerically_2019,ridley_lead_2019,krivenko_dynamics_2019,chen_auxiliary_2019,kleinhenz_dynamic_2020,atanasova_correlated_2020,Erpenbeck_Resolving_2021,erpenbeck_revealing_2021,kleinhenz_kondo_2022,erpenbeck_quantum_2023,Erpenbeck_Shaping_2023,atanasova_stark_2024,erpenbeck_steady_2024,Kunzel_Numerically_2024} in the hybridization expansion \cite{werner_continuous-time_2006}, for more details we refer the reader in particular to Refs.~\cite{Cohen_Taming_2015, erpenbeck_quantum_2023, erpenbeck_steady_2024}.
        The bath is parametrized by a wide, flat band with smooth cutoffs, 
        $\Gamma(\epsilon) = \Gamma/[(1+e^{\eta(\epsilon-\omega_c)})(1+e^{-\eta(\epsilon+\omega_c)})]$, with 
        $\Gamma=1$ which serves as our unit of energy, and
        $\eta=10/\Gamma$, $\omega_c=25\Gamma$.
        We consider an Anderson impurity model with the parameters $U=-2\epsilon_0=8\Gamma$, which makes the system particle-hole symmetric. 
        All restricted propagators presented in the following are converged with respect to the hybridization expansion order and the size of an inchworm step.

        Fig.~\ref{fig:propagators} shows QMC results alongside the corresponding ESPRIT extrapolations for the restricted propagator. In this analysis, ESPRIT is applied to extrapolate all four propagator components simultaneously, using a single set of shared exponentials and matrix-valued prefactors \cite{zhang_minimal_matrix_2024}.
        As the propagator decays to zero and its final value is of no interest, we do not apply any special treatment to account for an infinite value.
        Panel (a) presents results for the standard ESPRIT algorithm with all exponents included and no additional postprocessing, 
        while panel (b) shows results with exponentially increasing components removed (see Sec.~\ref{sec:post_processing} for details).
        In both panels, the top two rows show the restricted propagators of the empty/doubly occupied and spin-polarized impurity states, with the temperature decreasing from left to 
        right.
        While particle-hole symmetry reduces the number of physically inequivalent propagators to two which are explicitly shown in the figure, all four are retained since they are computed independently and differ slightly due to Monte Carlo noise.
        The spin-polarized propagators (middle rows) exhibit increasingly long-lived oscillations at lower temperatures, a hallmark of strong correlations. 
        To assess the quality of the extrapolation, we consider the maximum deviation between ESPRIT and QMC data. 
        We find that ESPRIT, with or without postprocessing of the exponents, reproduces the QMC data with a maximum error below $10^{-3}$ using only the short data segment up to the indicated sampling time $t_{\text{samp}}$ (red dashed line), which ranges from $0.3/\Gamma$ to $5/\Gamma$. The bright green lines show the corresponding final extrapolation associated with this $t_{\text{samp}}$.
        The bottom row in both panels of Fig.~\ref{fig:propagators} shows the shared exponents $\xi_p$ as a function of sampling time, with opacity scaled by the maximum absolute value of the corresponding prefactor $C_p$ to visually emphasize dominant contributions and aid the distinction between signal and noise.

        We first analyze the standard ESPRIT algorithm in Fig.~\ref{fig:propagators}(a), without applying any of the postprocessing steps outlined in Sec.~\ref{sec:post_processing}. 
        All exponents identified by ESPRIT, including exponentially increasing ones, are included, as evident from the extrapolations at very short sampling times (fine red lines in the top two rows), which exhibit diverging behavior. 
        Despite this, ESPRIT reliably extrapolates the full dynamics from the short data segment up to the indicated sampling time $t_{\text{samp}}$ (red dashed line), ranging from $0.5/\Gamma$ to $5/\Gamma$ depending on temperature, using only a small number of exponents ($M=4$--$9$). 
        This demonstrates ESPRIT's potential as an extrapolation tool for real-time quantum impurity solvers, providing compact representations with few poles and enabling long-time predictions from relatively short time dynamics. 
        However, extrapolations based on shorter time series are unstable, as illustrated by the thin red lines corresponding to truncated inputs at $0.2/\Gamma$, $0.5/\Gamma$, $1.0/\Gamma$, and $2.0/\Gamma$ (shown only when shorter than the required $t_{\text{samp}}$).

        Considering the results of ESPRIT with all increasing exponents removed, as detailed in Sec.~\ref{sec:post_processing} and shown in Fig.~\ref{fig:propagators}(b), the general observation that ESPRIT reliably extrapolates real-time dynamics remains valid. 
        With the diverging exponents excluded, no exponentially growing extrapolations are observed, even for very short sampling times, although small differences between the extrapolations and the actual data remain (see e.g. $\beta=1.0/\Gamma$). 
        Removing the exponentially increasing components allows ESPRIT to reliably predict the dynamics from shorter input data segments compared to standard ESPRIT that includes all exponents. 
        A caveat is that more exponents are required as compared to the standard ESPRIT. 
        However, only a few of these additional exponents contribute significantly, as reflected by their prefactors: \
        for $\beta=0.01/\Gamma$ and $\beta=0.1/\Gamma$, only 3 of 8 exponents have prefactors with an absolute value that is larger than $10^{-2}$,
        for $\beta=1.0/\Gamma$, only 4 of 10,
        and for $\beta=10.0/\Gamma$, only 5 of 30. 
        Exponents with smaller contributions typically decay rapidly and could be disregarded for a more compact representation (not done here), though this can reduce the accuracy of the fit for the short-time data.

        Having shown that ESPRIT-based extrapolation can accurately predict long-time behavior from short-time dynamics, the key question becomes how to determine when sufficient information has been captured—so that further propagation adds no value and extrapolation becomes the preferred approach.
        To address this, we compute the ESPRIT decomposition at each time step and track how the extracted exponents evolve with time. 
        The bottom rows in both panels of Fig.~\ref{fig:propagators} show the exponents $\xi_p$ as a function of sampling time. 
        To visually highlight the most significant exponents, the opacity was scaled by the corresponding prefactor $C_p$, enhancing the distinction between signal and noise.
        We observe that, regardless of whether exponentially increasing exponents are filtered out, the exponents quickly settle into plateaus, indicating that the essential information has been captured and further propagation provides no significant improvement.
        The onset of these plateaus also matches the $t_{\text{samp}}$ required to reproduce the QMC results with good accuracy, as discussed above.
        This observation motivates a practical algorithm: during real-time propagation, ESPRIT can be performed at each time step, and once the exponents stabilize, e.g. remain unchanged within a set threshold across subsequent time steps, propagation can be terminated, and the remainder of the dynamics extrapolated. 
        This strategy is broadly applicable, easy to implement in real-time solvers, and will be explored further in future work.
        
        We note that a similar analysis -- characterizing extrapolation quality based on extracted exponents -- can also be performed using other methods that yield exponentials from real-time data such as Prony's method, the Matrix Pencil approach, or DMD. 
        Hence, a complementary analysis for DMD is provided in App.~\ref{sec:append_DMD_MC}.

    \subsection{Extrapolating infinite time behavior and localization in the spin-boson model}\label{sec:localization}
        \begin{figure}
            \raggedright (a)\\
            \centering
            \includegraphics{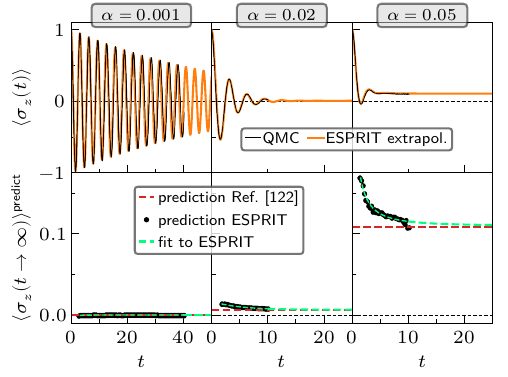}\\
            \raggedright (b)\\
            \centering
            \includegraphics{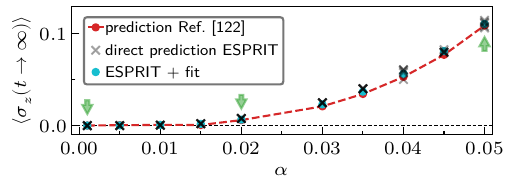}\\
            \caption{
                    Spin-polarization in the spin–boson model in the deep sub-Ohmic regime ($s=0.2$) at low temperature ($\beta=100/\Delta$).  
                    Panels (a), top: Time evolution of the spin-polarization for various coupling strengths $\alpha$; black lines are inchworm QMC data from Ref.~\cite{goulko_transient_2025}, orange lines are ESPRIT extrapolations.
                    Panels (a), bottom: Infinite-time spin-polarization. Red dashed lines are values from Ref.~\cite{goulko_transient_2025}, black dots show ESPRIT predictions from data up to time $t$, and the light green dashed line fits these to an exponential decay.  
                    Panel (b): Onset of localization as a function of $\alpha$. Red dashed lines are from Ref.~\cite{goulko_transient_2025}, gray crosses show ESPRIT predictions based on the full time series, successively truncated by removing up to the last five time steps, where many of these crosses are so close that they overlap in the plot and appear as black crosses.
                    Blue dots show the final value obtained by fitting an exponentially decaying function to the ESPRIT predictions over time (corresponding to light green dashed lines in panels (a), bottom).  
                    Green arrows indicate the data sets shown in panel (a).
                    }
            \label{fig:spin-boson}
        \end{figure}

        We now combine the two key capabilities of ESPRIT explored in the previous sections -- predicting long-time behavior and tracking the structure of the exponential representation over time -- to analyze real-time dynamics in the sub-Ohmic spin–boson model. Specifically, we apply ESPRIT to numerically exact inchworm QMC data for the spin-polarization following a quantum quench reported in Ref.~\cite{goulko_transient_2025}.
        In this recent work by some of us, the long-time limit of the spin-polarization was used to identify dynamical signatures of the zero-temperature quantum phase transition between localized and delocalized phases, with a nonzero steady-state polarization indicating localization, leading to a dynamical phase diagram that differs from its equilibrium counterpart.
        A key challenge is that delocalization can occur only after very long times, depending on the properties of the bosonic bath \cite{leggett_dynamics_1987}, which Ref.~\cite{goulko_transient_2025} addressed by fitting a physically motivated heuristic function to the finite-time data obtained from  inchworm QMC methods \cite{Chen_Inchworm_2017, Chen_Inchworm_2017_2, yang_inclusion_exclusion_2021, kim_pseudoparticle_2022, cai_fast_2022, cai_numerical_2023, Cai_Inchworm_2020, cai_bold_thin_bold_2023, wang_real_time_2023, wang_solving_2025, kim_vertex_based_2023}.
        Here, we show that ESPRIT enables a fully data-driven assessment of the dynamics, providing robust extrapolations and helping determine whether the available real-time data is sufficient. 
        This model- and ansatz-free approach extends the analysis in Ref.~\cite{goulko_transient_2025}.

        The spin–boson model is given by the Hamiltonian
        $H = H_\text{s} + H_\text{b} + H_\text{sb}$,
        with the spin term
        $H_\text{s} = \frac{\epsilon}{2}\sigma_z + \frac{\Delta}{2}\sigma_x$,
        the bosonic bath
        $H_\text{b} = \sum_l \hbar\omega_l b_l^\dagger b_l$,
        and the spin–bath coupling
        $H_\text{sb} = \frac{\sigma_z}{2} \sum_l c_l \left(b_l^\dagger + b_l\right)$.
        Here, $\sigma_i$ are Pauli matrices, $\epsilon$ is half the energy splitting between the two spin states, and $\Delta$ is the tunneling amplitude. The operators $b_l^{(\dagger)}$ are bosonic creation and annihilation-operators for mode $l$ with frequency $\omega_l$, and $c_l$ is the coupling between the spin and mode $l$.
        The sub-Ohmic bosonic bath is characterized by the spectral density
        $J(\omega) \sim \alpha \omega^s$,
        where $\alpha$ sets the overall coupling strength and $s<1$ defines the sub-Ohmic regime.
        The observable of interest in Ref.~\cite{goulko_transient_2025} is the spin-polarization at time $t$, $\braket{\sigma_z(t)}$. 
        Following a quantum quench from an initially decoupled spin and bosonic bath, 
        $\braket{\sigma_z(t)}$ is computed using the numerically exact real-time inchworm QMC method in the spin–bath coupling expansion \cite{Cohen_Taming_2015, Chen_Inchworm_2017, Chen_Inchworm_2017_2}. 
        The data we analyze from Ref.~\cite{goulko_transient_2025} are for $\epsilon = 0$, with energies measured in units of $\Delta$. 
        We focus on the data for the deep sub-Ohmic regime with $s = 0.2$ and low temperature, $\beta = 100/\Delta$.
        
        Fig.~\ref{fig:spin-boson}(a), top panels, show the time evolution of $\braket{\sigma_z(t)}$ computed using inchworm QMC for representative coupling strengths $\alpha$, alongside ESPRIT extrapolations that accurately capture the dynamics and extend them to longer times. 
        We note that the ESPRIT scheme used here discards all exponentially increasing contribution and sets the smallest absolute-value exponent to zero, which yields the most reliable extrapolation given the low noise in the QMC data.
        An alternative strategy, adding a zero exponent, provides inferior performance in the present case and requires longer-time data for robust results (not shown).
        The number of exponents $M$ was determined by setting the tolerance for accepting singular values in the ESPRIT algorithm to $10^{-3}$. The resulting $M$ depends on the data considered and ranges from $M = 3$ for small $\alpha \leq 0.02$ to $M = 23$ for $\alpha = 0.05$.
        The bottom panels of Fig.~\ref{fig:spin-boson}(a) plot the predicted infinite-time spin-polarization, $\braket{\sigma_z(t \rightarrow \infty)}^{\text{predict}}$ as a function of the maximum time $t$ used in the ESPRIT input.
        For weak coupling ($\alpha = 0.001$), ESPRIT consistently predicts zero polarization even with minimal data, matching results from Ref.~\cite{goulko_transient_2025}. 
        For $\alpha = 0.02$, and more notably for $\alpha = 0.05$, the prediction evolves with increasing time, indicating that while ESPRIT detects localized behavior, longer time series would improve accuracy.
        %Because ESPRIT predictions are relatively smooth functions of the available time 
        Because the predicted infinite-time value varies gradually with the data available to ESPRIT, with only minor changes as additional time points are included
        (due to low noise in the QMC data), we enhance the analysis by fitting an exponentially decaying function to $\braket{\sigma_z(t \rightarrow \infty)}^{\text{predict}}$ versus $t$ (shown as the bright green dashed line). This prediction from this fit aligns closely with the values reported in Ref.~\cite{goulko_transient_2025}. We also highlight that ESPRIT reliably distinguishes small non-zero values from zero, as evident for $\alpha = 0.02$.

        A more comprehensive overview is provided in Fig.~\ref{fig:spin-boson}(b), which compares the infinite-time spin-polarization from Ref.~\cite{goulko_transient_2025} -- obtained using a heuristic ansatz fitted to the dynamics -- with ESPRIT-based predictions. 
        Gray crosses show the direct ESPRIT results using the full time series, each computed with one to five of the final time steps removed, providing an estimate of the uncertainty for the method. 
        Blue dots represent the outcome when an exponentially decaying function is fitted to the ESPRIT predictions as a function of input time.
        Overall, we find excellent agreement between ESPRIT and the results of Ref.~\cite{goulko_transient_2025}, with the fit of an exponentially decaying function to the ESPRIT output yielding slightly closer alignment than the direct ESPRIT predictions.
        This agreement strongly supports the findings in Ref.~\cite{goulko_transient_2025}, which relied on fitting a heuristic functional form to the data, which could have been critiqued for the chosen ansatz and the potential for associated artifacts, as well as the sensitivity to initial conditions in nonlinear fitting routines.
        %In contrast, ESPRIT is a purely data-driven method that is agnostic to the underlying system and makes no assumptions about the functional form of the signal. 
        In contrast, ESPRIT is a purely data-driven method that is agnostic to the underlying system, as it does not require prior knowledge of the governing equations or the physical origin of the signal. It assumes only that the signal can be expressed as a sum of complex exponentials (with numerical feasibility determined by the number of terms required for an accurate representation). This exponential basis is flexible and well suited to capturing oscillatory and decaying behavior, which enables ESPRIT to be applied across a wide range of physics and engineering problems (though we note that some functions cannot be feasibly represented in this way).
        Its agreement with the heuristic results in Ref.~\cite{goulko_transient_2025} thus provides independent cross-validation and extends their analysis by extracting maximal information from the available data. 
        Moreover, by analyzing the functional form of the infinite-time prediction as a function of input duration, ESPRIT reveals when and how much additional real-time data would enhance the reliability of the long-time extrapolation.
        We view this as a compelling demonstration of ESPRIT's power and versatility in extracting long-time behavior from short-time dynamics 
        -- an ability with broad relevance to many real-time simulation and experimental contexts.
        However, obtaining an unambiguous prediction for the infinite-time limit is generally challenging and depends on both the system and the quality of the available data, making comparisons to cases where the infinite-time value is known (such as the equilibrium situation for the system studied here) a valuable complement to monitoring how the predicted infinite-time value evolves with the available input data.

\section{Summary and outlook}\label{sec:summary}

    Reliable extrapolation, denoising, and representation of real-time data can significantly enhance the quantitative analysis of quantum systems in both experimental and theoretical settings.
    In this work, we investigated the application of the ESPRIT algorithm to real-time data, representing dynamics as a sum of complex exponentials to enable analysis and reliable extrapolation of long-time behavior from short-time input.

    We introduced the ESPRIT methodology along with postprocessing techniques that improve robustness to noise and facilitate the extraction of specific features, such as the infinite-time limit.
    Benchmarking ESPRIT against an analytic test function, we compared its performance to other standard extrapolation methods and assessed its stability under noisy conditions. These tests demonstrate ESPRIT's ability to extract meaningful long-time behavior even in the presence of moderate noise.

    We then applied ESPRIT to two distinct observables obtained from QMC simulations of different physical systems.
    In the Anderson impurity model, ESPRIT enabled the extrapolation of restricted propagators deep into the correlated regime. A stability-based criterion on the extracted exponents identified when the short-time data fully capture the system's dynamics, allowing extrapolation to replace propagation.
    In a second application, ESPRIT accurately recovered the long-time spin polarization in the spin-boson model, capturing localization effects relevant to a quantum phase transition.

    Overall, our results establish ESPRIT as a robust and versatile tool for representing real-time data and extracting long-time behavior from short-time dynamics, with broad applicability in both experimental and computational physics. 
    A particularly promising direction is its integration with real-time impurity solvers, where it could substantially reduce the computational cost of studying strongly correlated systems with long coherence times. 
    While this work showcased applications within QMC frameworks, extending the approach to other widely used methods, such as time-dependent DMRG \cite{Cazalilla_Time_2002, White_Real_2004, Heidrich_Real_2009, Karrasch_Finite_2012, Paeckel_Time_2019} or TD-DFT \cite{Runge_Density_1984, Goings_Real_2018, Li_Real_2020}, presents a particularly promising direction for future research.
    Also, applying ESPRIT to experimental data, especially in regimes where traditional extrapolation and representation methods struggle, is another important next step, including testing its predictive power in noisy environments where tailored denoising and postprocessing could enhance accuracy.
    Further avenues include extending ESPRIT to regimes where the amount of available data is insufficient to reliably extract exponents and longer time series would be required, together with estimating the required data length or noise thresholds for stable predictions.
    Analyzing the structure of the extracted exponentials may also yield physical insight and reveal connections between different observables within the same system. 
    Finally, exploring ESPRIT for two-time objects, which form the basis of many theoretical approaches to quantum dynamics, offers a promising direction. 
    In particular, ESPRIT's minimal-pole representation could be used to compress two-time data, enabling nonequilibrium Green's function calculations over longer time intervals, as pursued in other compression and representation schemes \cite{kaye_low_2021, yin_using_2022, Meirinhos_Adaptive_2022, blommel_adaptive_2024, sroda_high_2024}.

\section{Acknowledgements}
    We thank Thomas Blommel, Martin Eckstein, Fabian K\"unzel, Dolev Goldberger, Hsing-Ta Chen, and Wei-Ting Lin for fruitful discussions. 
    This work was supported by the U.S.\ Department of Energy, Office of Science, Office of Advanced Scientific Computing Research and Office of Basic Energy Sciences, Scientific Discovery through Advanced Computing (SciDAC) program under Award No. DE-SC0022088. 
    L.Z. was supported by the National Science Foundation under Grant No. NSF QIS 2310182 and Y.Y. by the National Science Foundation under Grant No. NSF DMR 2401159.
    G.C.\ acknowledges support by the ISF (Grant No. 2902/21), by the PAZY Foundation (Grant No. 318/78) and by MOST NSF-BSF (Grant No. 2023720).
    O.G.\ is supported by the NSF under Grants No.~ PHY-2441282 , PHY-2112738 and OSI-2328774.

\appendix

    \section{Fourier transformation and ESPRIT} \label{sec:app_FT}
        As ESPRIT decomposes a given signal into a sum of complex exponentials, one major advantage is that the Fourier transform with respect to the zero time of a signal,
        \[
        f(\omega) = \int_0^\infty f(t) e^{i\omega t} \, dt,
        \]
        can be calculated analytically:
        \begin{eqnarray*}
            \mathrm{signal}\ f(t)
            \quad &\stackrel{\mathrm{ESPRIT}}{\longrightarrow}& \quad
            f(t) \approx \sum_{p=1}^M C_p e^{\xi_p t} \\
            \quad &\stackrel{\mathrm{FT}}{\longrightarrow}& \quad
            \tilde{f}(\omega) \approx \sum_{p=1}^M -\frac{C_p}{\xi_p + i\omega}.
        \end{eqnarray*}
        Since Fourier transforming data to analyze its frequency content or, equivalently, the peak structure of the spectrum is a common task in physics, we illustrate the performance of ESPRIT for this purpose using two test cases, one with a discrete spectrum and one with a continuous spectrum, each examined both with and without noise.
    
\begin{figure}[h!]
            \raggedright (a)\\
            \centering
            \includegraphics{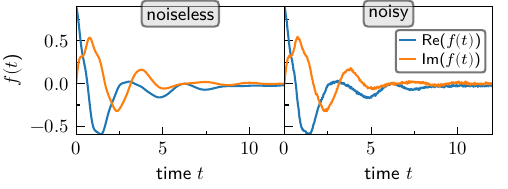}
            \raggedright (b)\\
            \centering
            \includegraphics{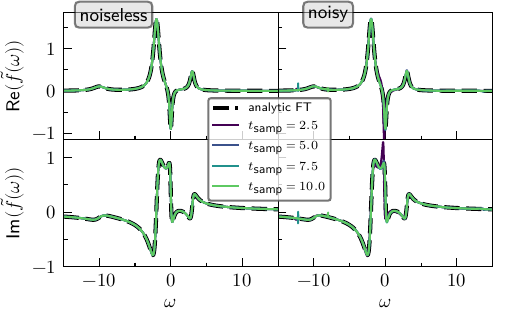}
            \raggedright (c)\\
            \centering
            \includegraphics{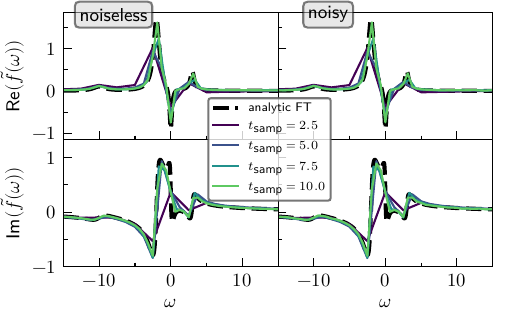}
            \caption{
                Illustration of the function $f(t)$, which consists of a finite number of exponential poles (Eq.~(\ref{eq:test_FT_1})), and its Fourier transform under different noise conditions.  
                Panles (a) Real and imaginary parts of the function $f(t)$ defined in Eq.~(\ref{eq:test_FT_1}) as a function of time. Noiseless data is shown in the left panel, and data with Gaussian noise with $\sigma = 10^{-2}$ is shown in the right panel.  
                Fourier transform of the same function computed with ESPRIT (panel b) and FFT (panel c), shown for noiseless (left panels) and noisy (right panels) cases.
                For both algorithms, the sampling time $t_{\mathrm{samp}}$ indicates the time up to which data is available to the algorithms.
                }
            \label{fig:FT1}
        \end{figure}

        To illustrate how accurately ESPRIT reproduces the peak structure in the Fourier transform of a function consisting of a discrete number of exponential poles, we compare its output to an analytic test function for which the exact Fourier transform is known. We use the analytic example,
        \begin{eqnarray}
            f(t) &=& c_1 e^{-o_1 t} + c_2 e^{-o_2 t} + c_3 e^{-o_3 t} + c_4 e^{-o_4 t},
            \label{eq:test_FT_1}
        \end{eqnarray}
        with parameters $c_1 = 0.85$, $c_2 = 0.15$, $c_3 = 0.1$, $c_4 = -0.2$ and
        $o_1 = \frac{1}{2} - 2i$, $o_2 = \frac{1}{3} + 3i$, $o_3 = 1 - 10i$, and $o_4 = 0.2$.
        Note that this function is equivalent to the analytic function from Sec.~\ref{sec:noisy_test}, subtracting the constant term that would otherwise introduce a singularity at zero frequency. 
        Fig.~\ref{fig:FT1}(a) shows the real and imaginary parts of $f(t)$ with and without noise.
        
        Fig.~\ref{fig:FT1}(b) shows the Fourier transform obtained with ESPRIT, both with and without noise, for data available up to a finite sampling time $t_{\mathrm{samp}}$. The results are compared to the exact analytic Fourier transform. ESPRIT achieves excellent accuracy even when provided with data over relatively short time intervals. In the noisy case, some spurious spikes appear, particularly at short $t_{\mathrm{samp}}$. Increasing $t_{\mathrm{samp}}$ leads to a more stable and accurate reconstruction.
        
        For comparison, Fig.~\ref{fig:FT1}(c) shows results obtained with a standard FFT. While FFT can also reproduce the Fourier transform, it generally requires a larger $t_{\mathrm{samp}}$ to reach a comparable level of accuracy and tends to have difficulty resolving closely spaced features in frequency space when only limited time-domain data is available. Increasing $t_{\mathrm{samp}}$ mitigates these issues to a certain extend.
    
\begin{figure}[h!]
            \raggedright (a)\\
            \centering
            \includegraphics{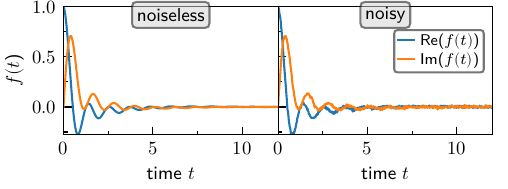}
            \raggedright (b)\\
            \centering
            \includegraphics{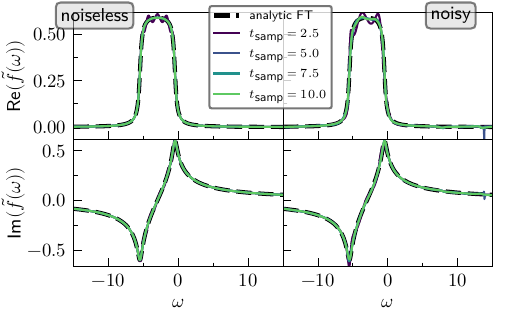}
            \raggedright (c)\\
            \centering
            \includegraphics{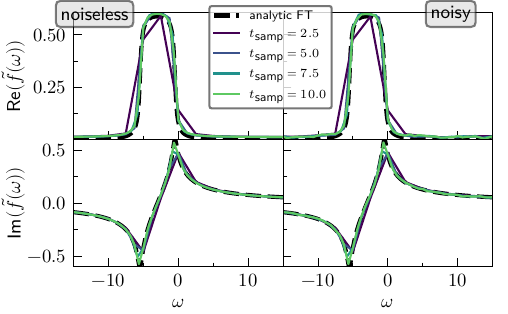}
            \caption{
                Illustration of the function $f(t)$, which consists of a continuum of exponential poles (Eq.~(\ref{eq:test_FT_2}), )and its Fourier transform under different noise conditions.  
                Panels (a): Real and imaginary parts of the function $f(t)$ defined in Eq.~(\ref{eq:test_FT_2}) as a function of time. Noiseless data is shown in the left panel, and data with Gaussian noise with $\sigma = 10^{-2}$ is shown in the right panel.  
                Fourier transform of the same function computed with ESPRIT (panel b) and FFT (panel c), shown for noiseless (left panels) and noisy (right panels) cases. 
                For both algorithms, the sampling time $t_{\mathrm{samp}}$ indicates the time up to which data is available to the algorithms.
                }
            \label{fig:FT2}
        \end{figure}

        We now repeat the analysis of ESPRIT's accuracy in reproducing the Fourier transform, this time for a function composed of a continuous distribution of exponential poles. Since ESPRIT assumes that the signal can be decomposed into a sum of complex exponentials, there are inherent limits to the types of functions it can accurately represent. For numerical efficiency, it is desirable that only a small number of exponentials suffices to describe the signal. Functions that are continuous superpositions of exponentials are, in principle, challenging, as an exact representation would require an impractically large number of terms.
        In practice, however, we find that even such functions can often be well approximated using only a small number of exponentials. To illustrate this, we consider a simple example consisting of a continuous superposition of exponentials,
        \begin{eqnarray}
            \hspace{-0.4cm}
            f(t)    &=& \int_{\xi_0}^{\xi_1} C\, e^{(-b + i\xi) t} \, d\xi
            = \frac{C e^{-bt}}{i t} \left( e^{i\xi_1 t} - e^{i\xi_0 t} \right),
        \label{eq:test_FT_2}
        \end{eqnarray}
        with parameters $b = 0.25$, $C = 0.2$, $\xi_0 = 0.5$, and $\xi_1 = 5.5$. 
        The $1/t$ prefactor makes this function challenging to represent accurately with only a few exponentials. 
        For the specific function $f(t)$ considered here, the Fourier transform admits a known closed-form expression, allowing it to be evaluated analytically.
        Fig.~\ref{fig:FT2}(a) shows the real and imaginary parts of $f(t)$ with and without noise.
        
        Fig.~\ref{fig:FT2}(b) shows the Fourier transform obtained with ESPRIT, both with and without noise, for data truncated at a finite sampling time $t_{\mathrm{samp}}$. 
        We restrict ESPRIT to $M = 5$ exponentials to demonstrate that even a small number of terms can capture the essential features of a function with a continuous spectrum. 
        The results are compared to the exact analytic Fourier transform. 
        As in the case of a discrete spectrum, ESPRIT achieves excellent accuracy even for short $t_{\mathrm{samp}}$. 
        For short time intervals, small oscillations appear on top of the continuous background, and in the noisy case, additional spurious spikes are observed. 
        Generally, increasing $t_{\mathrm{samp}}$ improves stability and accuracy.
        
        For comparison, Fig.~\ref{fig:FT2}(c) shows results from a standard FFT. 
        While FFT can also reproduce the Fourier transform, it generally requires a larger $t_{\mathrm{samp}}$ to achieve comparable accuracy. 
        Artifacts in the Fourier transform are progressively reduced as $t_{\mathrm{samp}}$ increases.

    \section{Technical Details on DMD and Comparison with ESPRIT} \label{sec:append_DMD}

        As there is a substantial DMD community, and because DMD shares certain similarities with ESPRIT, we provide some additional technical details on DMD and its relation to ESPRIT in this Appendix. 
        We also include some further comparisons to illustrate these connections. 
        We emphasize, however, that this is \textit{not} a DMD-focused paper, the method is included solely for context, and no effort was made to optimize its performance.

        \subsection{Equivalence of ESPRIT and Hankel-DMD Eigenvalues}\label{sec:append_DMD_ESPRIT_comnparison}

            ESPRIT and DMD share similarities since both methods can extract complex exponentials from a given dataset. To further demonstrate the connection between these two, we focus on a specific variant of DMD, namely, the Hankel-DMD method~\cite{arbabi2017ergodic}. 
            The Hankel-DMD can be formulated using the same Hankel matrix employed in ESPRIT, thereby placing the comparison on an equal footing.
            \\
    
            {\em (a) Setup and definitions --}
            Assume a dataset consisting of ``snapshots'' \(f_k \in \mathbb{C}\). For simplicity, we consider the case where the \(f_k\) are scalar complex values, consistent with our example in Fig.~\ref{fig:method_comparison}, but the extension to \(f_k \in \mathbb{C}^n\) is straightforward. 
            The dataset admits a decomposition into $M$ damped complex exponentials $\lambda_p =  e^{\xi_p \Delta t}$,
            \begin{equation}
            f_k = \sum_{p=1}^M C_p \, \lambda_p^{\,k} ,
            \end{equation}
            in line with Eq.~(\ref{eq:compact_exp}) of the main text. Here, $C_p$ are the amplitudes determining the contribution of each exponential to the overall signal, and $\Delta t$ is the spacing of the equidistant grid on which the signal is discretized (see main text).
            Defining the delay vectors at point $k$,
            \begin{equation}
            y_k = \begin{bmatrix} f_k \\ f_{k+1} \\ \vdots \\ f_{k+N-L} \end{bmatrix}
            \in \mathbb{C}^{N-L+1},
            \end{equation}
            Hankel matrices can be expressed as
            \begin{equation}
            H_0 = [y_0,\dots,y_L], \qquad
            H_1 = [y_1,\dots,y_{L+1}].
            \end{equation}
            Further, we define the ``steering'' vectors
            \begin{equation}
            s_i = \begin{bmatrix} C_i \\ C_i \lambda_i \\ \vdots \\ C_i \lambda_i^{N-L} \end{bmatrix},
            \qquad
            S = [s_1,\dots,s_M],
            \end{equation}
            and a Vandermonde-type coefficient matrix 
            $T \in \mathbb{C}^{M\times L+1}$ 
            with entries
            $T_{i,k} = \lambda_i^{\,k}$,
            where $i=1,2, \dots, M$ and $k=0,1,\dots, L$.
            Then
            \begin{equation}
            H_0 = S T, \qquad H_1 = S \Lambda T, 
            \label{eq:hankel_DMD1}
            \end{equation}
            where \(\Lambda = \mathrm{diag}(\lambda_1,\dots,\lambda_M)\).
            \\
            
            {\em (b) Hankel-DMD --}
            Following the outline of the DMD method from Sec.~\ref{sec:other_alogs}, we
            start by taking the rank-$r$ SVD 
            \begin{equation}
                H_0 = U_r \Sigma_r V_r^\dagger . \label{eq:Hankel_SVD_app}
            \end{equation}
            Because $\mathrm{range}(H_0) = \mathrm{range}(S)$,
            there exists an invertible matrix $K \in \mathbb{C}^{r\times r}$ such that
            \begin{equation}
            S = U_r K, \qquad T = K^{-1} \Sigma_r V_r^\dagger.
            \label{eq:hankel_DMD2}
            \end{equation}
            The reduced operator, defined in Eq.~(\ref{eq:reduced_op}) of the main text, is then given by
            \begin{equation}
            \tilde A = U_r^\dagger H_1 V_r \Sigma_r^{-1}.
            \end{equation}
            Using Eqs.~\eqref{eq:hankel_DMD1} and \eqref{eq:hankel_DMD2}, this expression can be rewritten as
            \begin{equation}
            \tilde A = U_r^\dagger (U_r K \Lambda K^{-1} \Sigma_r V_r^\dagger) V_r \Sigma_r^{-1}
            = K \Lambda K^{-1}.
            \end{equation}
            Thus Hankel-DMD produces $\tilde A$ similar to $\Lambda$, so its eigenvalues are $\{\lambda_p\}$. \
            The calculation of the expansion coefficients follows the standard DMD procedure by using \( \mathbf{b} = \Phi^+ \mathbf{x}_0 \), where \( \Phi = H_1V_r \Sigma_r^{-1} K \). 
            \\

            {\em (c) Relation to ESPRIT --}
            The algorithm, as outlined in Sec.~\ref{sec:esprit}, also begins with an SVD of the Hankel matrix, Eq.~\eqref{eq:Hankel_SVD_app}.
            We then define the block-row selection matrices
            \begin{equation}
            J_1 = [\,I_{N-L} \;\; 0\,], \qquad
            J_2 = [\,0 \;\; I_{N-L}\,],
            \end{equation}
            where $J_1$ selects the first $N-L$ rows of a matrix and $J_2$ the last $N-L$ rows.
            The matrices $U_0$ and $U_1$ defined in the main text can then be expressed as
            \begin{equation}
                U_0 = J_1 U_r, \qquad U_1 = J_2 U_r .   \label{eq:U0U1_defs}
            \end{equation}
            When applied to the steering matrix $S$, the selection matrices satisfy the shift-invariance property,
            \begin{equation}\label{eqn:shift_inv}
                J_2 S = J_1 S \Lambda.
            \end{equation}
            Using $U_r = S K^{-1}$, i.e. Eqs.~\eqref{eq:hankel_DMD2} and \eqref{eqn:shift_inv}, we obtain 
            \begin{eqnarray}
            J_2U_r =J_1U_r( K \Lambda K^{-1})  \label{eq:J2_identity}
            \end{eqnarray}            
            Using the definition of the rotation matrix from Eq.~\eqref{eq:rotation_matrix} and the notation introduced in Eq.~\eqref{eq:U0U1_defs}, together with the identity in Eq.~\eqref{eq:J2_identity}, we obtain the rotation matrix in the form
            \[
            \Phi = (J_1 U_r)^{+} J_1 U_r (K \Lambda K^{-1}),
            \]
            which shows that ESPRIT also identifies the eigenvalues $\{\lambda_p\}$.
            \\

            {\em (d) Noisy data --}
            As shown above, in the noiseless case, both Hankel-DMD and ESPRIT yield reduced matrices similar to \(\Lambda\), and therefore recover the same eigenvalues \(\{\lambda_p\}\) corresponding to the underlying complex exponentials. In practice, however, the data is inevitably contaminated by noise, leading to perturbed Hankel matrices \(H_0\) and \(H_1\). In such noisy cases, numerical implementations of the Hankel-DMD and ESPRIT algorithms generally produce different eigenvalues associated with the extracted exponentials. 
            In ESPRIT, the eigenvalues are extracted from the reduced matrices \(U_0\) and \(U_1\) by solving the 
            rotational-invariance relation \(U_1 = U_0 \Phi\). This explicitly enforces the underlying structure of the problem and tends to yield robust eigenvalue estimates in the presence of noise.
            Hankel-DMD, on the other hand, does not explicitly enforce the rotational-invariance relation and employs the reduced operator \(\tilde{A}\) (see Eq.~\eqref{eq:reduced_op}), which is the best-fit linear map between the delay-embedded snapshots in a least-squares sense.
            Moreover, differences in how the prefactors are determined further distinguish the results of the two methods. Generally speaking, it is easier to filter the unphysical exponents in ESPRIT without introducing global bias, since its prefactors are obtained by solving a linear system constructed from the recovered eigenvalues (see Eq.~\eqref{eq:Vandermonde}). In contrast, the prefactors in Hankel-DMD depend on \(\Phi^{+}\), where the DMD eigenmodes contained therein cannot be as easily adjusted with respect to the change of filtered eigenvalues. The difference in the numerical results will be illustrated later in Fig.~\ref{fig:propagators_Hankel_DMD}.
            \\

        \subsection{Comparing Hankel-DMD and ESPRIT for the analytic test function in Sec.~\ref{sec:noisy_test}}\label{sec:append_data_HankelDMD}
            \begin{figure}[!htbp]
                \raggedright (a)\\
                \centering
                \includegraphics{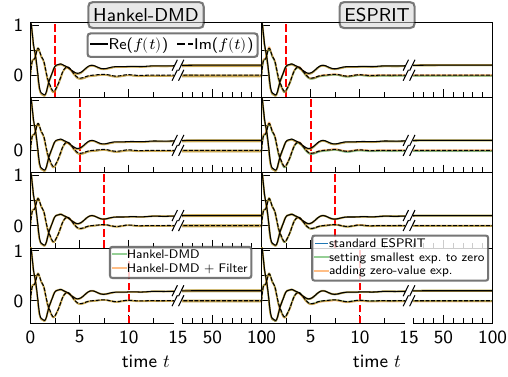}\\
                \raggedright (b)\\
                \centering
                \includegraphics{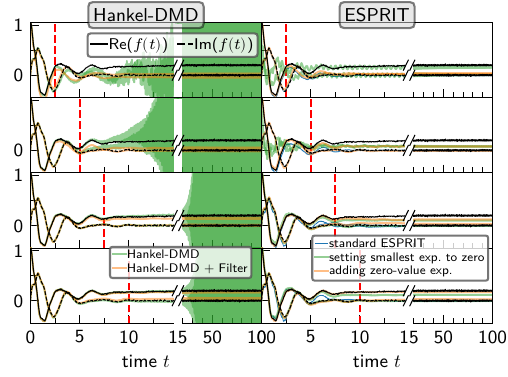}
                \caption{
                Direct comparison of Hankel-DMD, as introduced in this Appendix, and ESPRIT based on the extrapolation of the function \(f(t)\) [Eq.~(\ref{eq:test_func})].
                This figure complements Fig.~\ref{fig:method_comparison} from the main text.
                As in Fig.~\ref{fig:method_comparison}, the function \( f(t) \) is shown as solid and dashed black lines for the real and imaginary parts, respectively.
                Colored lines represent the results obtained with different versions of Hankel-DMD and ESPRIT.
                The label ``Hankel-DMD+Filter'' indicates that exponentially growing modes have been removed, as is also done for ESPRIT (see Sec.~\ref{sec:post_processing} for details).
                All numerical parameters shared by Hankel-DMD and ESPRIT are identical, the threshold in their SVD decompositions set to \(10^{-6}\), and both methods receive the same Hankel matrix as input.
                Shaded regions correspond to extrapolations with rapid oscillations, which cause the area to appear filled.
                Panel (a): without noise. 
                Panel (b): with Gaussian noise of standard deviation \(\sigma = 10^{-2}\).
                In both panels, the sampling time \( t_{\text{samp}} \) (red vertical dashed line), i.e.\ the time up to which data is available to the algorithms, increases from top to bottom, with \( t_{\text{samp}} = 2.5,\, 5.0,\, 7.5,\) and \(10.0\), respectively.
                }
                \label{fig:DMD_ESPRIT_Fig1}
            \end{figure}
            \begin{figure}[tb]
                \raggedright (a)\\
                \centering
                \vspace*{-0.25cm}
                \hspace{0.7cm}\textbf{Interpolation Error}\\
                \vspace*{+0.25cm}
                \includegraphics{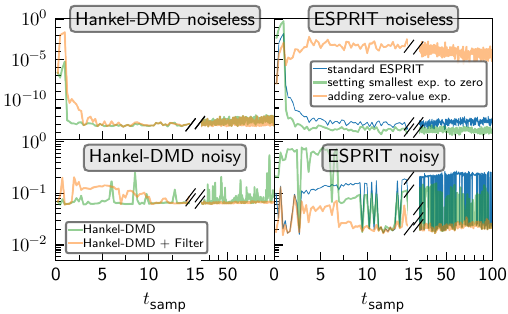}\\
                \vspace{0.25cm}
                \raggedright (b)\\
                \centering 
                \vspace*{-0.25cm}
                \hspace{0.7cm}\textbf{Extrapolation Error}\\                \vspace*{+0.25cm}
                \includegraphics{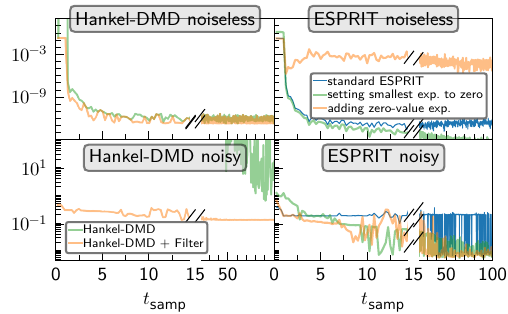}
                \caption{
                Comparison of the interpolation (panels (a)) and extrapolation (panels (b)) errors obtained with Hankel-DMD and ESPRIT as a function of the sampling time \( t_{\text{samp}} \).
                This analysis is based on the analytic function \( f(t) \) [Eq.~(\ref{eq:test_func})], both without noise and with Gaussian noise of standard deviation \( \sigma = 10^{-2} \).
                The error shown corresponds to \( \max_{t \in \text{grid}} | f(t) - f_{\text{rec}}(t) | \), 
                where \( f_{\text{rec}}(t) \) denotes the reconstructed function as obtained from Hankel-DMD and ESPRIT, respectively. 
                The error is evaluated on a uniform grid ranging from \( 0 \) to \( t_{\text{samp}} \) for interpolation (panels (a)) and from \( t_{\text{samp}} \) to \( 200 \) for extrapolation (panels (b)), with a grid spacing of \( 0.025 \).
                Colored lines represent the results obtained by different versions of Hankel-DMD and ESPRIT.
                The label ``Hankel-DMD+Filter'' indicates that exponentially growing modes have been removed, as is also done for ESPRIT (see Sec.~\ref{sec:post_processing} for details).
                All numerical parameters shared by Hankel-DMD and ESPRIT are identical, with the threshold in their SVD decompositions set to \( 10^{-6} \), and both methods receive the same Hankel matrix as input.
                }
                \label{fig:DMD_ESPRIT_Error}
            \end{figure}

            Given the connection between ESPRIT and DMD, particularly the Hankel-DMD variant introduced in the previous section, we provide a more detailed comparison of their performances in this Appendix, which is complementary to the analysis in Sec.~\ref{sec:noisy_test}.
            While the main text focuses on ESPRIT and includes other methods only in their most basic forms for context (as this work neither aims to compare methods comprehensively nor to emphasize DMD in particular),
            this Appendix offers a closer examination of Hankel-DMD, the DMD formulation most closely related to ESPRIT.
            We emphasize that, although additional details are provided here, the DMD implementation remains deliberately basic and does not include advanced enhancements sometimes employed in the DMD community, as this is not the scope of this work.
            
            Fig.~\ref{fig:DMD_ESPRIT_Fig1} presents a direct comparison between Hankel-DMD and ESPRIT, analogous to Fig.~\ref{fig:method_comparison} from the main text, based on the analytic test function \( f(t) \) defined in Eq.~(\ref{eq:test_func}) and used in Sec.~\ref{sec:noisy_test}.
            Panels~(a) show the case without noise, while panels~(b) include Gaussian noise with standard deviation \( \sigma = 10^{-2} \).
            The data available to the algorithms increase from top to bottom as indicated by red dashed lines.
            The ESPRIT variants shown are those discussed in the main text and include the postprocessing step that removes exponentially growing contributions, as well as two versions explicitly enforcing a finite final value (see Sec.~\ref{sec:post_processing} for details).
            For comparison, we include results from the basic Hankel-DMD implementation with all modes retained (green lines in Fig.~\ref{fig:DMD_ESPRIT_Fig1}), as well as a version in which exponentially growing modes are filtered out by setting the corresponding exponents to zero, since removing diverging modes by setting their prefactors to zero can introduce artifacts (see App.~\ref{sec:append_DMD_MC} and Fig.~\ref{fig:propagators_Hankel_DMD}). This approach is analogous to ESPRIT's postprocessing (orange lines in Fig.~\ref{fig:DMD_ESPRIT_Fig1}).  
            We do not include any explicit treatment of the finite asymptotic value for Hankel-DMD for the following reason. While ESPRIT naturally allows adding specific exponents prior to solving the Vandermonde system in Eq.~(\ref{eq:Vandermonde}), thereby fitting both the original and added modes to the data, DMD constructs extrapolations by projecting the data onto the computed dynamical modes. Although mode removal in DMD is possible (albeit potentially problematic, as exemplified in App.~\ref{sec:append_DMD_MC} and Fig.~\ref{fig:propagators_Hankel_DMD}), introducing new modes with desired properties is nontrivial and beyond the scope of this work. Setting the smallest DMD exponent to zero, in analogy to ESPRIT, is in principle possible; however, because the basic DMD variant used here does not update its modes based on the changed exponents, this procedure does not affect the reconstruction for the problem and timescales considered here.

            Comparing the noiseless results in Fig.~\ref{fig:DMD_ESPRIT_Fig1}(a), both methods yield excellent extrapolations even with limited data, for all ESPRIT variants as well as for Hankel-DMD, both with and without filtering.
            The situation changes in the noisy case in Fig.~\ref{fig:DMD_ESPRIT_Fig1}(b). While ESPRIT’s performance and its variants were discussed in detail in the main text, we find for Hankel-DMD that filtering out exponentially growing modes is essential for reliable predictions, which reach a level of accuracy comparable to ESPRIT on the scale shown in Fig.~\ref{fig:DMD_ESPRIT_Fig1}.

            A more detailed comparison of the performance of Hankel-DMD and the different ESPRIT variants considered in this work is shown in Fig.~\ref{fig:DMD_ESPRIT_Error}.
            Specifically, the figure displays the maximal absolute value difference between the reconstructed signal \( f_{\mathrm{rec}}(t) \) obtained from the various Hankel-DMD and ESPRIT implementations as a function of the sampling time \( t_{\text{samp}} \), evaluated on an equidistant time grid with resolution \(\Delta t = 0.025\). 
            All numerical parameters shared by Hankel-DMD and ESPRIT are identical, the threshold in their SVD decompositions set to \( 10^{-6} \), the criterion for filtering eigenvalues is identical and detailed in the main text, and both methods receive the same Hankel matrix as input.
            All errors are plotted on a logarithmic scale.
            Fig.~\ref{fig:DMD_ESPRIT_Error}(a) compares the interpolation accuracy, that is, how well the algorithms reproduce the data between \(0\) and \(t_{\text{samp}}\). 
            For the noiseless case (top row), all variants of Hankel-DMD and ESPRIT achieve extremely accurate interpolations for any \(t_{\text{samp}}\) that is not extremely short. 
            The only outlier is the ESPRIT variant with an additional zero component. Fitting the data with an explicitly included zero exponent alongside the original (numerically small but finite) component introduces a minor ambiguity in the fit, resulting in slightly larger errors on the order of \(10^{-5}\).
            For the noisy data (bottom row of Fig.~\ref{fig:DMD_ESPRIT_Error}(a)), both Hankel-DMD and ESPRIT again perform comparably, with errors of the order of the noise level. 
            For interpolation, including or excluding exponentially diverging modes does not produce a systematic difference for Hankel-DMD. 
            The standard ESPRIT occasionally alternates between two distinct error levels at longer sampling times, while the ESPRIT variant with an added zero exponent yields the lowest overall error which is roughly a factor of four to five smaller than that of Hankel-DMD.

            Fig.~\ref{fig:DMD_ESPRIT_Error}(b) compares the extrapolation accuracy of the different Hankel-DMD and ESPRIT variants, that is, how well the algorithms generalize and extend the data to times \(t \geq t_{\text{samp}}\). This question of extrapolation is more closely aligned with the scope of this work. 
            In the noiseless case (top row in Fig.~\ref{fig:DMD_ESPRIT_Error}(b)), all variants of Hankel-DMD and ESPRIT achieve extremely accurate extrapolations for any \(t_{\text{samp}}\) that is not extremely short. The only outlier is ESPRIT with an additional zero component, which shows a residual error of only \(10^{-4}\)–\(10^{-5}\). 
            The situation changes again in the presence of noise, as shown in the bottom row of Fig.~\ref{fig:DMD_ESPRIT_Error}(b). Hankel-DMD without exponential filtering produces errors so large that most data points lie outside the plotted range. Only at long times does this error re-enter the plot due to the finite grid size. Hankel-DMD with exponentially growing modes removed remains stable, with a typical error around \(2 \times 10^{-1}\). 
            While this is an improvement, this error is more than an order of magnitude larger than the noise level of \(10^{-2}\), and given that the long-time limit of the signal is \(f_\infty = 0.02\), the method does not allow to reliably predict the correct asymptotic value. 
            The same is true for standard ESPRIT, especially for short times, although at long times its error jumps repeatedly below this level. This behavior aligns with one of the motivations of the present work, which aims to correctly identify the long-time limits of the dynamics. Only the ESPRIT variants that set the smallest value to zero or add an additional zero component achieve lower errors, reaching at least the noise level at long times. 
            We note that the data shown here represent only a single example and a more detailed analysis is left for future work.

        \subsection{DMD-based extension of real-time data for propagation methods}\label{sec:append_DMD_MC}
            In this Appendix, we extend the analysis from Sec.~\ref{sec:MC_extension} to HO-DMD as well as Hankel-DMD, complementing our earlier comparison of extrapolation methods -- especially given that DMD has demonstrated success in systems with long coherence times \cite{yin_using_2022, Reeves_Dynamic_2023,kaneko_forecasting_2025}.
            
            \begin{figure}[!htbp]
                \raggedright (a)\hspace{3.6cm}{$\mathbf{n_s=1}$}\\
                \centering
                \includegraphics[width=0.46\textwidth]{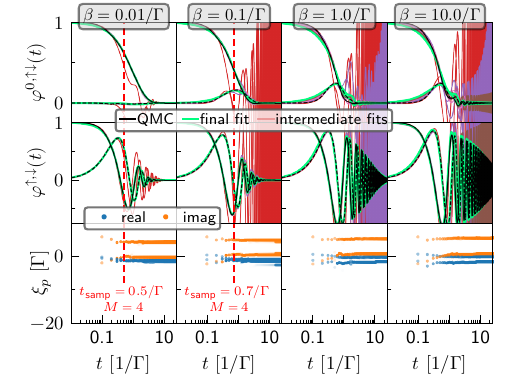}\\\vspace{0.05cm}
                \raggedright (b)\hspace{3.6cm}{$\mathbf{n_s=2}$}\\
                \centering
                \includegraphics[width=0.46\textwidth]{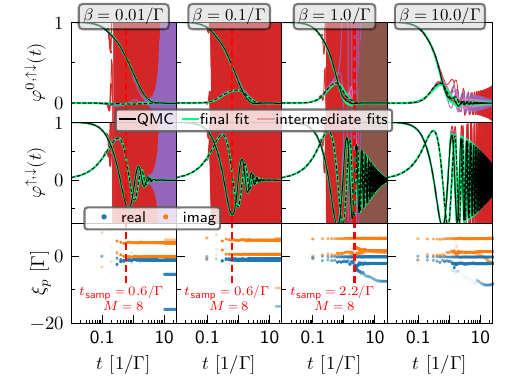}\\\vspace{0.05cm}
                \raggedright (c)\hspace{3.6cm}{$\mathbf{n_s=3}$}\\
                \centering
                \includegraphics[width=0.46\textwidth]{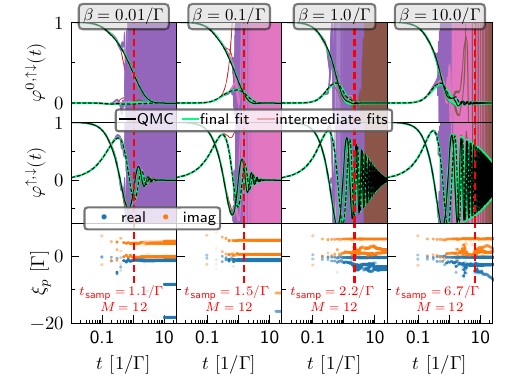}
                \caption{
                    Application of the HO-DMD algorithm to QMC data for the single-orbital Anderson impurity model across increasing inverse temperatures $\beta$ (left to right). 
                    Panel (a): $n_s=1$; Panel (b): $n_s=2$; Panel (c): $n_s=3$.
                    These plots are equivalent to Fig.~\ref{fig:propagators}(a) from the main text, but using HO-DMD instead of ESPRIT.
                    Shaded regions appear filled due to extrapolations with rapid oscillations.
                    Faint red lines indicate predictions for sampling times $0.2/\Gamma$, $0.5/\Gamma$, $1.0/\Gamma$, and $2.0/\Gamma$, where applicable.
                    The tolerance for the SVD decomposition is \(10^{-9}\).
                    The bottom row in all panels displays the extracted exponents, which are shared by all restricted propagator components.
                }
                \label{fig:propagators_DMD}
            \end{figure}
            \begin{figure}[!htbp]
                \raggedright (a) \hspace{0.8cm}{Hankel-DMD -- without filtering of exponents}\\
                \centering
                \includegraphics[width=0.46\textwidth]{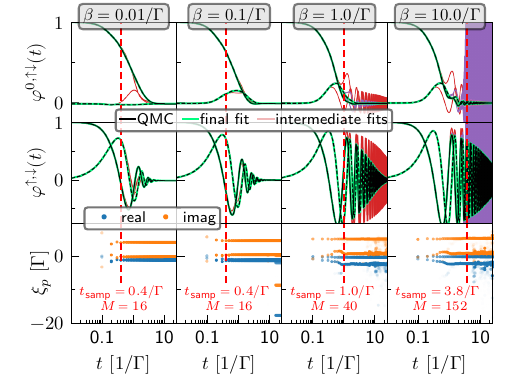}\\
                \vspace{0.05cm}
                \raggedright (b) \hspace{0.6cm}{Hankel-DMD -- increasing exponents set to zero}\\
                \centering
                \includegraphics[width=0.46\textwidth]{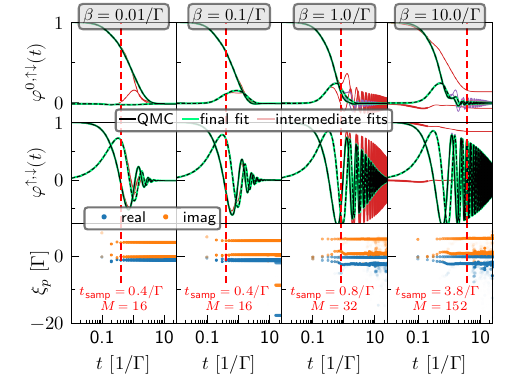}\\
                \vspace{0.05cm}
                \raggedright (c)  \hspace{0.cm}{Hankel-DMD -- prefactors of increasing exps.\ set to zero}\\
                \centering
                \includegraphics[width=0.46\textwidth]{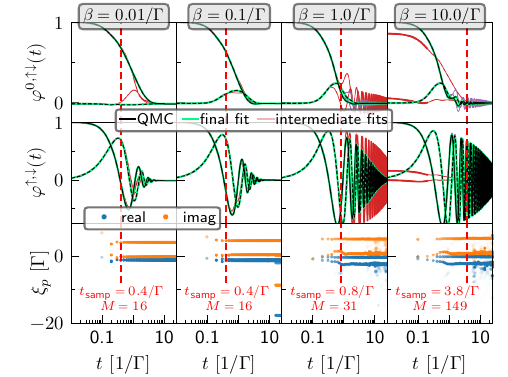}  
                \caption{
                    Application of the Hankel-DMD algorithm to QMC data, with inverse temperatures $\beta$ increasing from left to right. 
                    Panel (a): No postprocessing or exclusion of exponentials. 
                    Panels (b) and (c): With postprocessing where exponentially increasing exponents are either replaced by zero (b), or removed by setting the corresponding prefactor to zero (c). 
                    These plots correspond to Fig.~\ref{fig:propagators}(b) from the main text, but use Hankel-DMD instead of ESPRIT.
                    The tolerance for the SVD decomposition is \(10^{-9}\).
                    Faint red lines indicate predictions for sampling times $0.2/\Gamma$, $0.5/\Gamma$, $1.0/\Gamma$, and $2.0/\Gamma$, where applicable.
                    The bottom row in all panels displays the extracted exponents, which are shared by all restricted propagator components.
                }
                \label{fig:propagators_Hankel_DMD}
            \end{figure}
            
            Fig.~\ref{fig:propagators_DMD} repeats the analysis from Sec.~\ref{sec:MC_extension} for HO-DMD. 
            As with ESPRIT, HO-DMD is applied to extrapolate all four propagator components simultaneously, employing a single set of shared exponentials and matrix-valued prefactors.
            For a given $n_s$, HO-DMD thus employs up to $4 n_s$  exponents to represent the data, where the factor of $4$ reflects the four propagators included in the snapshot matrix. 
            Although particle–hole symmetry implies only two physically inequivalent propagators, all four are computed by QMC and therefore differ slightly due to MC noise, remaining distinct at the SVD threshold used here.
            The panels (a)--(c) correspond to different values of $n_s$, as described in Sec.~\ref{sec:other_alogs}. 
            This implementation represents the most basic and textbook form of DMD, without any postprocessing or removal of exponentially growing components, analogous to the unprocessed ESPRIT results shown in Fig.~\ref{fig:propagators}(a).
            All numerical parameters shared between ESPRIT and HO-DMD are identical, and the threshold for the SVD decomposition is set to $10^{-9}$, and the criterion for filtering eigenvalues is identical and detailed in the main text.
            As in Sec.~\ref{sec:MC_extension}, the vertical red dashed lines mark the sampling time $t_{\text{samp}}$ 
            at which a prediction accuracy of $10^{-3}$ with respect to the QMC data is achieved, ranging from $0.5/\Gamma$ to $6.7/\Gamma$ depending on temperature. 
            The bright green lines indicate the extrapolations associated with these $t_{\text{samp}}$ values. 
            $M$ denotes the number of exponentials, consistent with its usage for ESPRIT.
            If no red dashed line is shown for a given $n_s$-value and temperature, the target accuracy was not reached even when using the full data set; in that case, the bright green line corresponds to the interpolation based on the full data.
            Overall, the HO-DMD results in Fig.~\ref{fig:propagators_DMD} resemble those obtained with ESPRIT in Sec.~\ref{sec:MC_extension}, demonstrating that HO-DMD can successfully perform the same type of analysis and yield comparable results.
            However, compared to standard ESPRIT, HO-DMD appears somewhat less stable when only short-time data is available, as it appears more prone to exponentially increasing components. 
            Also, HO-DMD requires longer time-series data, as reflected by the larger $t_{\text{samp}}$ values compared to those in Fig.~\ref{fig:propagators}(a), 
            and more exponential terms to achieve similar precision.
            For $n_s = 1$, representations with the desired accuracy are obtained only for the two highest temperatures as the algorithm only has $4$ exponents to represent the data. 
            As $n_s$ increases, so does the number of exponentials, and representations achieving the target accuracy also become available for lower temperatures. 
            However, this comes at the cost of requiring larger $t_{\text{samp}}$ values and more exponents, and occasionally leads to the introduction of highly suppressed exponential terms when longer-time data is used, likely to better capture short-time features at the expense of the long-time prediction.
            The tendency of HO-DMD to require more exponents and longer sampling times may stem from the fact that DMD-based approaches do not exploit the rotational invariance property inherent to ESPRIT, which enables ESPRIT to identify dominant exponentials more efficiently in noisy settings such as the one studied here. 
            Postprocessing steps, such as filtering out exponentially increasing components, may improve the performance of HO-DMD; we apply these steps to Hankel-DMD for a direct comparison with ESPRIT in the next paragraph. 
            Also, more advanced DMD variants and strategies addressing these issues likely exist, but as this is not a DMD-focused study, we do not explore them further here.

            Fig.~\ref{fig:propagators_Hankel_DMD} repeats the analysis from Sec.~\ref{sec:MC_extension} for Hankel-DMD, applied to all four propagator components simultaneously to extract a single set of shared exponentials and matrix-valued prefactors.
            As outlined at the beginning of this Appendix, this version of DMD shares the most similarities with ESPRIT and
            all numerical parameters shared between ESPRIT and Hankel-DMD are identical for this comparison, the threshold for the SVD decomposition is again set to $10^{-9}$, the criterion for filtering eigenvalues is identical and detailed in the main text, and both methods receive the identical Hankel matrix as input.
            To allow for an objective comparison, we apply the same postprocessing procedures as used for ESPRIT. 
            Fig.~\ref{fig:propagators_Hankel_DMD}(a) shows results without any postprocessing, directly corresponding to Fig.~\ref{fig:propagators}(a), where no exponents were filtered in ESPRIT. 
            Panels (b) and (c) address exponentially increasing contributions, analogous to the treatment in Fig.~\ref{fig:propagators}(b). 
            In ESPRIT, divergent exponents are filtered by removing them before solving the Vandermonde system in Eq.~(\ref{eq:Vandermonde}) and refitting the remaining components to the data. 
            Hankel-DMD, by contrast, does not employ a Vandermonde system but projects the data onto dynamic modes. 
            For the most direct comparison, we implemented two analogous filtering schemes: 
            in Fig.~\ref{fig:propagators_Hankel_DMD}(b), exponentially increasing exponents were set to zero, while in Fig.~\ref{fig:propagators_Hankel_DMD}(c), the corresponding modes were removed by setting their prefactors to zero. 
            We note that more sophisticated approaches to handle such cases may exist within the DMD literature, but these are beyond the scope of this work, as our goal here is to provide a straightforward and transparent comparison with ESPRIT.

            Overall, the Hankel-DMD results in Fig.~\ref{fig:propagators_Hankel_DMD} closely resemble those obtained with ESPRIT in Sec.~\ref{sec:MC_extension}, and even more so than the HO-DMD results in Fig.~\ref{fig:propagators_DMD}, demonstrating that Hankel-DMD can successfully perform the same type of analysis with comparable accuracy.
            When compared directly with HO-DMD, Hankel-DMD achieves the target precision of $10^{-3}$ for all data sets, independent of whether exponentially increasing contributions are filtered, and requires shorter sampling times $t_{\text{samp}}$, albeit at the cost of more exponentials, similar to ESPRIT.
            Comparing Hankel-DMD without filtering to its ESPRIT counterpart in Fig.~\ref{fig:propagators}(a), both methods reach similar accuracies, but Hankel-DMD does so with slightly shorter $t_{\text{samp}}$, while ESPRIT requires roughly four times fewer exponentials.
            Filtering out exponentially increasing exponents in Hankel-DMD mainly removes the divergence at very short $t_{\text{samp}}$, with only minor effects on the number of exponents $M$ and the sampling time needed to reach $10^{-3}$ accuracy. 
            Setting diverging exponents to zero replaces their contributions with constant terms, so that they still count toward $M$, whereas removing them entirely by setting their prefactors to zero reduces $M$ slightly. 
            Both strategies have drawbacks, particularly at the lowest temperature $\beta = 10.0/\Gamma$ and short sampling times.
            Setting the diverging exponents to zero effectively turns their contributions into a constant offset, causing the extrapolated signal to plateau at long times. In contrast, removing these exponents entirely eliminates components essential for accurately capturing the short-time dynamics, thereby worsening the short-time fit.
            These issues diminish for larger $t_{\text{samp}}$.
            Unlike Hankel-DMD, ESPRIT avoids these artifacts by refitting all retained exponents to the data via the Vandermonde condition after filtering. Consequently, ESPRIT with exponentially increasing exponents removed yields a more compact representation and achieves the same accuracy with fewer exponents and shorter sampling times.
            Finally, when comparing the extracted exponents (bottom rows of Fig.~\ref{fig:propagators_Hankel_DMD} and Fig.~\ref{fig:propagators}(b)), ESPRIT produces more stable results with respect to $t_{\text{samp}}$, especially at low temperatures, whereas Hankel-DMD tends to introduce numerous weakly contributing exponents (visible as faint blue dots with negative values). This likely stems from ESPRIT's use of rotational invariance, which enhances its robustness to noise. Further improvements to DMD-based approaches might be achieved by using a more detailed treatment of dynamical modes, by rescaling modes that are kept rather than removing them, or adopting more advanced variants, which are beyond the scope of this work.

            To summarize, we note that DMD-methods also exhibit a plateau-like stabilization of the extracted exponents beyond a certain time, suggesting that, as with ESPRIT, a criterion can be established to determine when all relevant information is already contained in the short-time dynamics and further time propagation can be replaced by extrapolation.

    \section{Machine learning approach for dynamics and long-time predictions}\label{sec:append_ML}
        This Appendix provides additional details on the machine learning approach used to predict the system's behavior and its long-time limit.

    \subsection{Details of the dynamics prediction }\label{sec:append_ML_dyn}
        The RNN architecture used in this work to predict the dynamics of the data shown in Fig.~\ref{fig:method_comparison} consists of two LSTM layers followed by a multilayer perceptron (MLP) that maps the hidden states to the network output~\cite{zhu_predicting_2025}. 
        Each LSTM layer has a hidden size of $64$. For complex-valued data, the real and imaginary parts are treated as separate input channels. Thus, for the analytic function \( f(t) \) considered in Sec.~\ref{sec:noisy_test}, the input is a two-dimensional vector \([\,\mathrm{Re}\,f(t_i),\, \mathrm{Im}\,f(t_i)\,]\) for \(1 \le i \le N_t\), where \(N_t\) denotes the length of the training dataset. The corresponding output consists of the predicted time derivatives \([\,\mathrm{Re}\,f'(t_i),\, \mathrm{Im}\,f'(t_i)\,]\). 
        The predicted dynamics is then obtained by solving the differential equation for the observable using the RNN-predicted derivatives, i.e.,~\(\text{RNN}: f(t)\mapsto f'(t).\)

    \subsection{Details of the infinite-time limit  prediction }\label{sec:append_ML_inf}
        \begin{figure}[tb]
            \raggedright (a) \hspace{0.45\linewidth}(b)\\[0.5em]
            \raggedright
            \includegraphics{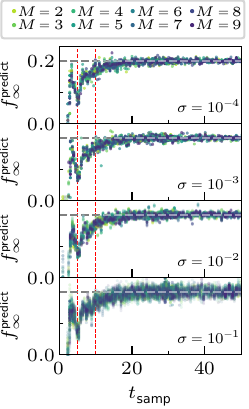}\\
            \vspace{-5.5cm}
            \hspace{5cm}
            \includegraphics[scale=0.4]{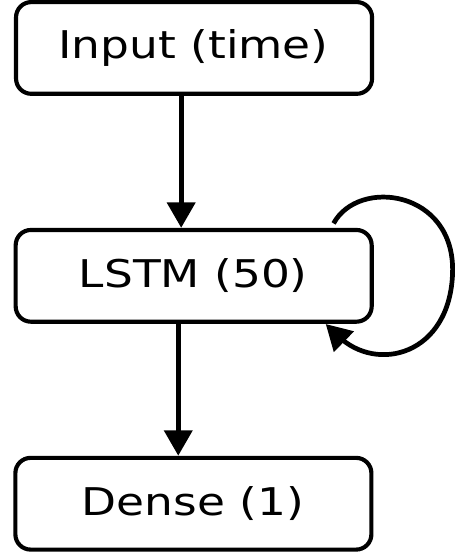}\\
            \vspace{1.7cm}
            \caption{
                    Influence of noise on the fidelity of predicting the infinite-time value using a recurrent neural network (RNN). 
                    Panel (a): RNN-based counterpart of Fig.~\ref{fig:noise_analysis}(a) of the main text:
                    The dot colors indicate the fixed number of exponential terms $M$  used in training the network. 
                    The red dashed lines highlight the values $t_{\text{samp}}=5.0$ and $t_{\text{samp}}=10.0$ and serve as a guide for the eye and indicate the regime of $t_{\text{samp}}$ for which an accurate extrapolation is most relevant.
                    Panel (b): schematic representation of the RNN used for this analysis.
                    }
            \label{fig:ML_final_val}
        \end{figure}

        Sec.~\ref{sec:noisy_test} featured a RNN-based approach to extrapolate a given dataset. 
        Rather than predicting the entire dynamics from a limited time-series, neural networks can also be trained to specifically predict the infinite-time value. 
        This can complement the analysis in Sec.~\ref{sec:ESPRIT_noise_analysis} by using a neural network in place of ESPRIT. To provide a more comprehensive picture, we showcase the result of this strategy in this Appendix.
        
        To estimate only the infinite-time value $f(t\rightarrow\infty)$ rather than the full dynamics, we employed a neural network consisting of a single hidden LSTM layer with $50$ units \cite{Hochreiter1997}, followed by a dense output layer with a single linear activation, which is schematically shown in Fig.~\ref{fig:ML_final_val}(b). 
        The network was trained using the Adam optimizer \cite{kingma2017adammethodstochasticoptimization} and a mean squared error (MSE) loss function. 
        Training was performed on $10{,}000$ unique samples, presented in batches of $100$.
        Each training sample was generated from
        $f(t) = \sum_{p=1}^{M} C_p e^{\xi_p t}$,
        with $C_p$ drawn uniformly from $[-1, 1]$. 
        For $p = 1, \dots, M-1$, the real parts of $\xi_p$ were sampled from $[-5, -0.002]$ and the imaginary parts from $[-1, 1]$, while $\xi_M = 1$ ensured a non-zero asymptotic value. 
        Gaussian noise with standard deviation $\sigma$ was added to study the influence of noise. 
        The resulting training data closely resembles the characteristics of our test function Eq.~(\ref{eq:test_func}).  
        We trained the network on functions with a fixed number of exponential terms $M$, reflecting the ESPRIT analysis where the number of exponents was also fixed. 
        The test function with noise was sampled on an equidistant time grid over the interval $[0, t_{\text{samp}}]$ with time step $\Delta t = 0.01$, consistent with the setup in Eq.~(\ref{eq:test_func}), which uses five exponentials.
        We emphasize that the model used here is deliberately simple, to provide a reference point for what is achievable with basic machine learning methods.

        Fig.~\ref{fig:ML_final_val}(a) shows the RNN-predicted asymptotic value $f_\infty^{\text{predict}}$ as a function of $t_{\text{samp}}$, analogous to the ESPRIT-based analysis in Fig.~\ref{fig:noise_analysis}(a). 
        Different colors indicate different numbers of exponentials $M$ used for training the RNN, with the color coding being equivalent to the one used in Fig.~\ref{fig:noise_analysis}(a).
        The results demonstrate that even a simple RNN can reliably predict the long-time limit, provided that a sufficiently long time series is provided. 
        For short $t_{\text{samp}}$, the RNN underestimates $f_\infty$, similar to the qualitative behavior observed with ESPRIT. 
        Across all four panels, corresponding to different noise levels, the prediction accuracy appears largely insensitive to both the noise level and the number of exponentials used for training.
        Comparing to the ESPRIT results in Fig.~\ref{fig:noise_analysis}(a), we find that ESPRIT generally achieves more accurate predictions and requires shorter $t_{\text{samp}}$, particularly in low to intermediate noise regimes. 
        However, in the high noise case, the RNN performs comparably to slightly better, suggesting potential advantages of neural network approaches in noisy environments.
        While ESPRIT outperforms the simple RNN used here under low noise conditions, more advanced and problem-specific RNN architectures could offer enhanced performance.

%\clearpage
\bibliography{bib}
\end{document}